
\input macnum.tex
\input preprint.sty

\def\psib{\overline\psi}
\def\chib{\overline\chi}

\def\lsim{\raise0.3ex\hbox{$<$\kern-0.75em\raise-1.1ex\hbox{$\sim$}}}
\def\gsim{\raise0.3ex\hbox{$>$\kern-0.75em\raise-1.1ex\hbox{$\sim$}}}
%

\font\its=cmti10
%
\def\s#1{\tilde{{\bf #1}}}

\def\MEF{m_{\rm eff}}\def\mef{\ifmmode\MEF\else$\MEF$\fi}

\def\Tr{{\rm Tr}}
\def\ie{{i.e.\/}}

\def\LMS{\Lambda_{\rm{\bar{M}\bar{S}}}}
\def\lms{\ifmmode\LMS\else$\LMS~$\fi}
\def\TC{T_c/\lms}\def\tc{\ifmmode\TC\else$\TC~$\fi}
\def\STR{\sqrt{\sigma}/\lms}\def\str{\ifmmode\STR\else$\STR~$\fi}
\def\NT{N_{\tau}}\def\nt{\ifmmode\NT\else$\NT~$\fi}
\def\NTA{N_{\tau}=8}\def\nta{\ifmmode\NTA\else$\NTA~$\fi}
\def\NTB{N_{\tau}=16}\def\ntb{\ifmmode\NTB\else$\NTB~$\fi}
\def\MEV{{\rm MeV}}\def\mev{\ifmmode\MEV\else$\MEV~$\fi}
\def\NS{N_{\sigma}}\def\ns{\ifmmode\NS\else$\NS~$\fi}
\def\LAT{N_{\tau} \times N_{\sigma}^3}\def\lat{\ifmmode\LAT\else$\LAT~$\fi}
%

%

\def\NP{{\sl Nucl.\ Phys.\ }}
\def\PL{{\sl Phys.\ Lett.\ }}
\def\PR{{\sl Phys.\ Rev.\ }}

\def\PRL{{\sl Phys.\ Rev.\ Lett.\ }}

\def\ZP{{\sl Z.\ Phys.\ }}

\def\banner{\hfill\hbox{\vbox{\offinterlineskip
                              \binum
                              \hlrznum}}\relax}
\def\manner{\hbox{\vbox{\offinterlineskip
                        \binum\hlrznum\date
           }}\hfill\relax}
\footline={\ifnum\pageno=1\manner\else\hfil\number\pageno\hfil\fi}
{\vsize=18truecm\banner\bigskip\baselineskip=15pt

\bigskip\bigskip\bigskip\bigskip\bigskip
\begingroup\titlefont\obeylines
\hfil Numerical Simulations in Particle Physics\hfil
\endgroup\bigskip
\centerline{
  Frithjof Karsch{$^{1,2}$}
  and Edwin Laermann{$^{2}$} }
\bigskip
\centerline{
$^1$~HLRZ, c/o KFA J\"ulich, D-5170 J\"ulich, Germany}
\centerline{
$^{~~2}$~Fakult\"at f\"ur Physik, Universit\"at Bielefeld,
        D-4800 Bielefeld 1, Germany}

\vskip 3.5truecm

\centerline{\bf ABSTRACT}\medskip
Numerical simulations have become an important tool to understand
and predict non-perturbative phenomena in particle physics. In this
article we attempt to present a general overview over the field.
First, the basic concepts of lattice gauge theories are described,
including a discussion of currently used algorithms and the
reconstruction of continuum physics from lattice results. We then
proceed to present some results for QCD, both at low energies
and at high temperatures, as well as for the electro-weak sector of
the standard model.
\vskip 2cm\noindent
- submitted for publication to ''Reports on Progress in Physics" -

}
\vfill\eject

\contents
\bigskip
\entry{1.}{Introduction}{3}
\entry{2.}{Particle Physics on the Lattice}{5}
\subentry{2.1.}{Lattice Regularized Quantum Field Theory}{5}
\subentry{2.2.}{Numerical Simulations}{10}
\subentry{2.3.}{Continuum Limit and Critical Phenomena}{18}
\entry{3.}{QCD at Low Energies}{22}
\subentry{3.1.}{Confinement and the Heavy Quark Potential}{23}
\subentry{3.2.}{Chiral Symmetry}{26}
\subentry{3.3.}{Meson and Baryon Spectroscopy}{27}
\subentry{3.4.}{Glueball Spectroscopy}{29}
\subentry{3.5.}{Weak Matrix Elements}{30}
\entry{4.}{QCD at High Temperature}{33}
\subentry{4.1.}{A New Phase of Matter}{33}
\subentry{4.2.}{Deconfinement}{34}
\subentry{4.3.}{The Chiral Phase Transition}{36}
\subentry{4.4.}{Spatial Correlations in the Quark-Gluon Plasma}{39}
\entry{5.}{Electro-Weak Sector of the Standard Model}{41}
\subentry{5.1.}{Pure Higgs Systems}{42}
\subentry{5.2.}{The Higgs Phase Transition}{43}
\subentry{5.3.}{Higgs-Yukawa Couplings}{45}
\subentry{5.4.}{Strong-Coupling QED}{48}
\entry{6.}{Outlook}{50}
\entry{}{Acknowledgments}{51}
\entry{}{References}{52}
\entry{}{Figure Captions}{60}
\vfill\eject

\section{Introduction}

There is a general belief among particle physicists that the fundamental
forces in physics are described by Quantum Field Theories (QFT's) with
local gauge invariance.
Although, at present, there does not exist a satisfactory framework for the
inclusion of the gravitational force, the remaining
electromagnetic, weak and strong
forces have been combined in a unified model -- the standard model.
The standard model is based on three symmetry groups, $U(1)$ and
$SU(2)$ for electromagnetism and weak isospin and colour
$SU(3)$. These theories have been tested experimentally and at
least in the perturbative regime the theoretical predictions have been
found to be in remarkable agreement with experiment.
However, these theories also incorporate
many important non-perturbative concepts, like confinement,
spontaneous chiral symmetry breaking or the Higgs mechanism. These properties
lead to drastic consequences for the spectrum of these
theories as well as for their phase structure at non-zero temperature.
The quantitative
study of such non-perturbative aspects of a QFT, e.g. a determination of the
mass spectrum or the location of a phase transition, in a perturbative context
is rather limited and non-perturbative approaches are needed to make
any progress
in a quantitative analysis of the consequences of these crucial
features of a QFT.

The exploration of field theories and the study of their
non-perturbative structure
received new impetus in the 70's through a newly developed
regularization scheme for the singularities of QFT's -- the
lattice regularization.
The formulation of lattice regularized field theories (Wilson 1974)
has put the
connection between the Euclidean path integral formulation of QFT's
and statistical
models on 4-dimensional lattices on a solid footing. It immediately
opened the possibility for applying standard methods from
statistical physics, which
have been used there since many years, to the analysis of QFT's.

Naturally these new lattice models first have been
studied using the well developed
high and low temperature series expansion techniques known in statistical
physics (Drouffe and Itzykson 1978). However, it was an obvious
step also to attempt to apply
Monte Carlo simulation techniques (Binder 1979) to the rather
complex field theoretic
models, once the computers were powerful enough to lead to
statistically significant
results within a reasonable time scale. Indeed the first results
from such numerical studies were exciting (Creutz 1980 and for a
collection of early results see also
Rebbi 1983)  and
led to the hope that long standing
non-perturbative questions, like the confinement problem in QCD,
could be answered
quantitatively this way. In fact, these first studies paved the way for a new
branch in theoretical high energy physics -- lattice field theory.

Numerical studies are nowadays used to explore
the structure of almost all quantum
field theories of interest, ranging from 2-dimensional spin models to the
basic building blocks of our standard model for the fundamental forces in
nature (Creutz 1992).
In particular in the analysis of Quantum Chromo Dynamics (QCD)
much progress has been
made towards a quantitative determination of such fundamental properties of the
theory as its particle spectrum, the strength of the confinement force and the
phase structure at finite temperature, although the required computer resources
became much larger than originally anticipated. In fact, today several groups
working in the field decided to develop dedicated machines for their
projects (Marinari 1992) and there even exist advanced
plans for the construction of
massively parallel, dedicated computers with
Teraflops performance (Christ 1990 and Negele 1992).
Consequently a good fraction
of research activities in the context of lattice field theories today is spent
in the development of new simulation techniques.
Substantial progress has been made
here for purely bosonic field theories, where cluster and multigrid algorithms
led to much improved simulation techniques. Crucial for the simulation of the
standard model, however, is the handling of fermions. Finding faster algorithms
for the fermionic sector of a QFT will be vital for future progress in this
field (Herrmann and Karsch 1991, Creutz 1992).

In this paper we will describe some of the basic algorithms
used for the simulation of
lattice field theories and review basic results that emerged from these
simulations. Although our notation is oriented towards
applications in the analysis of QCD, it applies, with minor modifications, to
other sectors of the standard model.
We will discuss in the next section
fundamental concepts in the formulation of lattice
field theories, putting emphasis on the currently used simulation
techniques and the
reconstruction of continuum physics from lattice calculations.
In the following sections
some basic results are discussed. In these sections we will,
where ever possible, try to avoid going
into a description of technical details of the actual Monte Carlo simulations.
Sections three and four are devoted to a presentation of results
obtained from simulations of QCD. In section three
we discuss results on the low
energy physics, while section four deals with the finite
temperature phase transition and
the physics of the quark-gluon plasma.
Similarly we discuss in section five
the zero and finite temperature physics of the electro-weak sector
of the standard model on the lattice.
Finally we give an outlook on future developments in
section 6.

\section{Particle Physics on the Lattice} \ssf
\subsection{Lattice Regularized Quantum Field Theory}

Almost all numerical studies of Quantum Field Theories (QFT's) are based
on the Euclidean path integral formulation of these
theories. The
vacuum to vacuum transition amplitude is given as a path integral over
all Euclidean field configurations
$$
Z_E = \int {\cal D} \Phi \, e^{-S_E} ~~,\EQNO{ze}
$$
where $S_E$ is the Euclidean action,
which defines the QFT in terms of a 4-dimensional
integral over the Lagrangian, ${\cal L}$,
$$
S_E = \int d^4x \, {\cal L}(\Phi)~~. \EQNO{ac}
$$
The Lagrangian depends only on the
fundamental fields $\Phi (t, \vec x)$ and a set
of coupling constants. Information on the spectrum, as well as various other
properties of the theory, can be obtained from the
behaviour of $n$-point Green functions,
$$
G(x_1,x_2,...,x_n) = Z_E^{-1} \int {\cal D}\Phi
\, \Phi(x_1)\Phi(x_2)...\Phi(x_n) \,
e^{-S_E}~~.\EQNO{green}
$$
For example, the mass $m_{\Phi}$ of a particle
created by the field $\Phi$ can be computed from the
exponential decrease of the two-point function at large time
separations $t$,
$$
\langle \Phi(t) \Phi(0) \rangle \;
{\buildrel{t \rightarrow \infty}\over \longrightarrow} \;
Z_{\Phi} \exp( - m_{\Phi} t)~~,
\EQNO{eqlattmass}
$$
where $Z_{\Phi}$ is the usual wave-function renormalization.

It is quite straightforward to generalize the path integral formulation of
a QFT to finite temperature (and chemical potential) (Bernard 1974).
The temperature
and volume of a thermodynamic system enter through the restriction of the
fundamental fields to a finite (3+1)-dimensional region of space-time.
In particular,
the temperature, $T$, enters by
restricting the Euclidean time interval to the range
$t \in [0,1/T]$ and by demanding periodic (anti-periodic) boundary
conditions for bosonic (fermionic) fields in this direction,
$$
Z(V,T) =
\int {\cal D} \Phi \, e^{-S_E(V,T )} ~~,\EQNO{partition}
$$
with
$$
S_E(V,T) = \int_0^{1/T} dx_0 \int_V d^3x {\cal L}(\Phi)~~. \EQNO{action}
$$
Thermodynamic quantities can then be obtained as derivatives of the
partition function $Z(V,T)$. For instance, the energy density and
the pressure are given by
$$
\epsilon = {T^2 \over V} {\partial \over \partial T} \ln Z\quad~~,~~\quad
p=T {\partial \over \partial V} \ln Z~~.~~
\EQNO{thermo}
$$

A QFT, defined formally by the above relations, needs to be regularized
in order to give a meaning to the path-integral and
to remove ultra-violet divergencies in the theory. This can be done by
introducing a cut-off in coordinate (or momentum) space.
A renormalization scheme, which allows to remove the
cut-off again, can then be defined and allows to separate divergent parts from
the physical, finite sector. For the purpose of numerical studies of QFT's
the introduction of a discrete space-time lattice is the most convenient
regularization scheme. It reduces the number of degrees of freedom to a large
but finite set and gives a well defined statistical interpretation to
the path integral and most observables of interest, which can be viewed as
expectation values calculated in a statistical ensemble with
Boltzmann weights $\exp(-S_E)$.

On a 4-dimensional space-time lattice with a lattice spacing ``$a$",
the fields $\Phi (x)$ are restricted to the discrete set of points,
$(x_0,\vec x) \rightarrow na \equiv (n_0a,n_1a,n_2a, n_3a)$.
Accordingly, $\Phi (x)$ gets replaced by $\phi(n)$
and the measure in the path integral becomes,
${\cal D} \Phi \rightarrow \prod_n
{\rm d}\phi (n)$. A finite, 4-dimensional lattice of size $\nt \times \ns^3$
does then represent a statistical system at temperature $T=1/\nt a$ in
a volume of size $V = (\ns a)^3$. The partition function of this system
reads
$$
Z_E = \int \prod_n {\rm d} \phi (n) \, e^{-S_E(V,T)} ~~.\EQNO{partlat}
$$
The crucial step in formulating a lattice regularized QFT is the proper
discretization of the Euclidian action, $S_E$. This can be achieved in a
straightforward way for a scalar field theory by discretizing the integral in
\eq{action} and replacing derivatives of fields by finite differences. The
action of the $\phi^4$-theory for instance, may be discretized as
$$\eqalign{
& \int {\rm d}^4x \,
\{{1\over 2}(\partial \Phi(x))^2 +{1\over 2}m^2_0 \Phi^2(x) +
{g_0\over 4 !}\Phi^4(x) \} \cr
\rightarrow
& \sum_{n=(n_0,..,n_3)} \bigl\{ -2 \kappa \sum_{\mu=0}^3
\phi (n) \phi (n+\hat{\mu}) +
\lambda [\phi^2(n) - 1]^2 + \phi^2(n) \bigr\}~~. \cr} ~~\EQNO{scalar}
$$
with $\hat{\mu}$ denoting the unit vector pointing to neighbouring sites in a
4-dimensional lattice and $\kappa, ~\lambda$ are couplings, which in the
naive continuum limit, $a \rightarrow 0$,
are related to $m_0 , g_0$ as $ g_0 = 6 \lambda / \kappa^2 ,
m_0^2 = {{1 - 2\lambda - 8 \kappa}\over{\kappa}}$.

In the case of a gauge theory the discretization is not at all so obvious. In
fact, it is important to choose a discretization such that the basic symmetries
of
the continuum action are preserved. This is not always possible as we will see
from our discussion of fermionic theories. However, the most important
step clearly is to construct a discretized action which preserves local
gauge invariance.
Wilson has shown first how one can achieve this (Wilson 1974).

\subsubsection{Gauge Fields on the Lattice}

Gauge fields mediate the interactions between matter. It is thus
suggestive
to introduce them as variables on the
links $(n,\mu)$ of a hypercubic, 4-dimensional lattice rather than
on the sites of the lattice.
Gauge fields, ${\cal A}_{\mu} (x)$, can
then be related to elements $U_\mu (n)$ of a gauge group. To be
specific we will consider the case of an $SU(N)$ gauge theory, which is
relevant for the formulation of QCD as well as the $SU(2)$-Higgs
model.\footnote{$^{\dagger}$}{The following discussion, of course, can
easily be reformulated for an abelian gauge theory like $U(1)$.}
On a lattice with lattice spacing $a$ the formal relation between
$U_\mu(n)$ and ${\cal A}_\mu (x)$ is given by
$$
U_\mu(n) =\exp\biggl[ -iga\sum_{b=1}^{N^2-1} \lambda^b {\cal A}_{\mu}^b
(x) \biggr ]~~,~~\EQNO{gauge}
$$
where $g$ is the bare coupling constant and $\lambda^b$ are the
generators of the group. Using this relation one can easily verify
that the single plaquette action proposed by Wilson
$$
S_G = {2N  \over g^2} \sum_{ n;   0 \le \mu < \nu \le 3 }
P_{\mu \nu}(n)
{}~,~~\EQNO{sg}
$$
with
$$
P_{\mu \nu}(n) = 1 - {1
\over N} Re \Tr U_\mu(n) U_\nu(n +\hat{\mu})
U_\mu^{-1}(n+ \hat{\nu})
U_\nu^{-1}(n)
{}~,~~\EQNO{plaq}
$$
approximates the continuum Lagrangian for the gauge fields up to
terms of $O(a^6)$,
${2N \over g^2} P_{\mu \nu}(n)
= {1 \over 4} a^4 F_{\mu \nu}^b F_{\mu \nu}^b +O(a^6)$.
In the continuum limit, $a \rightarrow 0$, these higher order
corrections become irrelevant.

\subsubsection{Fermions on the Lattice}

Wilson also suggested a discretization scheme for fermionic actions.
While it is easy to preserve local gauge invariance also in this case,
it is difficult to preserve all the chiral properties of fermionic actions.
Fermion actions contain only first
derivatives of the fields. As a consequence,
a straightforward
discretization, similar to the scalar case described in \eq{scalar},
leads to additional poles in the
lattice fermion propagator. In the continuum limit these additional poles
will give rise to 15 additional, unwanted fermion species rather than
only the one we started with.
It could be shown (Nielsen and Ninomiya 1981) that
this is a general phenomenon when in addition to
such elementary assumptions as
locality, hermiticity and translational invariance
also a continuous chiral symmetry of the
action is required.
There are however
certain loopholes.
Wilson proposed a discretization scheme for
fermions, in which
a second-order derivative term $S_W$
is added to the naively discretized fermion action $S_F$,
$$
\eqalign{
S_{WF} & =   S_F + S_W  \cr
       & =   {1\over 2} \sum_{x,\mu}
                \{ \psib(x) \gamma_{\mu} \psi(x+\hat{\mu})
                 - \psib(x+\hat{\mu}) \gamma_{\mu} \psi(x) \}  \cr
       & +    {1\over 2} \sum_{x,\mu}
                \{ 2 \psib(x) \psi(x)
                   - \psib(x+\hat{\mu}) \psi(x)
                   - \psib(x) \psi(x+\hat{\mu}) \}~~.\cr}
\EQNO{eqwilsonferm}
$$
The additional term $S_F$ becomes irrelevant in the (naive)
continuum limit.
Its effect is that the 15 additional fermions acquire a
large mass of $O(1/a)$, which diverges in the continuum limit,
and thus would
decouple from the dynamics of the theory. However, chiral
invariance of the action is lost at finite lattice spacing
and is to be recovered in the continuum limit.

Another approach, which we want to discuss in the following is
due to Kogut and Susskind (Kogut and Susskind 1975).
By distributing the four components of the continuum spinor over
different sites of the lattice it is possible to reduce the
number of additional species.
If one introduces $\bar{f}$ different staggered fermion species on the lattice
the staggered lattice action will lead to
$n_f =4 \bar{f}$ species of fermions in the continuum limit.
Moreover it preserves a global $U(\bar{f}) \times U(\bar{f})$ chiral
symmetry, i.e. an abelian subgroup of the continuum chiral symmetry.
For studies of chiral symmetry breaking as well as for finite temperature
calculations on the lattice it is convenient to
work with such a lattice action which preserves at least part of the
$SU(n_f) \times SU(n_f)$ chiral symmetry of the continuum action.
The staggered fermion action, obtained after a diagonalization
in the Dirac indices, becomes
$$
S_F= \sum_{i=1}^{\bar{f}} \sum_{n,m} \bar{\chi}_i(n) Q^i(n, m) \chi_i (m)
{}~.~~\EQNO{sf}
$$
Here the fermion fields, $\chi$, $\bar{\chi}$, are anticommuting
Grassmann variables defined on the sites of the lattice
and the fermion matrix $Q^i (n,m)$ is given by
$$
Q^i(n, m) = \sum_{\mu=0}^3 D_\mu(n, m) +m_i \delta (n,m)
{}~.~~\EQNO{matrix}
$$
The hopping matrices $D_{\mu} (n,m)$ mediate the nearest neighbour
interactions and
have non-zero elements only for
$m=n \pm \hat{\mu}$,
$$
D_\mu(n, m) = {1 \over 2} \eta_{\mu} (n) [ U_\mu(n) \delta
(n, m-\hat{\mu}) - U_\mu^{-1}(n) \delta(n,m +\hat{\mu})
]~~.     \EQNO{hopping}
$$
The phase factors $\eta_{\mu} (n) =
(-1)^{n_0+...+n_{\mu-1}}$ for $\mu > 0$ and $\eta_0 (n)= 1$
are remnants of the $\gamma_\mu$ matrices.
Finally the partition function takes on the form
$$
Z= \int \prod_{n,\mu} dU_\mu(n) \prod_{n,i} d\chi_i (n)
d\bar{\chi}_i(n) \, e^{\bigl [ -S_G -S_F \bigr]} ~~.~~\EQNO{partxx}
$$
As the fermionic part of the action is bilinear in the fields
$\bar{\chi}_i(n),~\chi_i(n)$, these can be integrated out and
the partition
function can be represented in terms of bosonic degrees of freedom only,
$$
Z= \int \prod_{n,\mu} dU_{n,\mu} \prod_{i=1}^{\bar{f}} detQ^i
e^ { -S_G } ~~.~~\EQNO{partsg}
$$
In this form the partition function is well suited for numerical studies. A
major problem is, however, caused by the presence of the fermion determinant,
which in general cannot be calculated exactly. Algorithms for the numerical
integration are thus required, which circumvent the explicit calculation
of this determinant.

\subsection{Numerical Simulations}
\subsubsection{Basic Concepts: Metropolis and Heat Bath Algorithm}

In the lattice regularization, the Feynman path integral,
\eq{partlat},
becomes a well-defined meaning as an
ordinary integral.
Because of the high-dimensionality, its
numerical evaluation, however, is a formidable task.
Imagine a lattice of just $10^4$ lattice
points, then \eq{partlat} represents a $10^4$ fold integral times
the number of internal degrees of freedom. Many field
configurations $\{ \phi\}$ will contribute to the integral
with rather small Boltzmann weights, $\exp\{ - S(\phi)\}$, though.
Thus an efficient
way to compute the integral would consist in generating a sequence
of field configurations $\{ \phi\}^{(k)}$ which are distributed
according to this weight factor. The
expectation value of an observable ${\cal O}(\phi)$
can then be approximated by the ensemble average
$$
\langle {\cal O} (\phi) \rangle =
{1\over M} \sum^M_{k=1} {\cal O} (\phi)^{(k)}~~.
\EQNO{edwin}
$$
Such a series of field configurations
is obtained by means of so-called Markov chains.
Starting from some arbitrary
initial configuration $\{\phi(n)\}^{(0)}$ one generates, one after the other,
new sets of $\phi$ fields. Under certain conditions, the sets
$\{\phi\}^{(k)}$ will be distributed according to
the equilibrium probability $\exp\{-S(\phi)\}$, once a number
of not-yet equilibrated initial configurations has been discarded.

The prototype algorithm which meets these conditions is the Metropolis
algorithm (Metropolis \etal 1953). It
consists of two steps: site by site (i) choose
a trial update $\phi'$ according to
some normalized probability distribution
$P_{trial}(\phi \rightarrow \phi') =
P_{trial}(\phi' \rightarrow \phi)$
and (ii) accept $\phi'$ with
the conditional probability
$$
P_{accept} = \min\left\{ 1,
{{e^{-S(\phi')}}\over {e^{-S(\phi)}}} \right\}~~.
\EQNO{eqaccept}
$$
The trial distribution $P_{trial}$ must be chosen
in such a way that the
whole configuration space can be covered.
The conditional accept probability $P_{accept}$ allows
for configurations
with a smaller Boltzmann weight
to be included in the set. This is necessary in order to account for
the quantum fluctuations.
Finally, the algorithm satisfies
detailed balance, $( P = P_{trial} * P_{accept} )$
$$
e^{-S(\phi)} P(\phi \rightarrow \phi') =
e^{-S(\phi')} P(\phi' \rightarrow \phi)~~,
\EQNO{eqbalance}
$$
which is a sufficient condition
for convergence to the equilibrium distribution.

As new configurations are calculated from previous ones, it is
clear that subsequent ``snapshots" of the system are not
statistically independent of each other. In order to
carry out a correct statistical error analysis it is
therefore desirable to step through configuration space rather quickly,
minimizing the number of intermediate configurations which have to
be discarded because they do not provide information independent of
the previous state.
The Metropolis algorithm is local and can be implemented efficiently.
However, either the new value $\phi'$ is close to the old
one, in which case the change in the action is small and its acceptance
is likely, or the new $\phi'$ is far from
the old one. In the latter case the change in the action is large, however,
and the acceptance rate drops exponentially. Both choices
result in a slow exploration of configuration space.
A somewhat better approach in this respect
has therefore been to pick a $\phi'(n)$ not randomly from the
whole manifold but with a weighting proportional to the Boltzmann
factor (Creutz \etal 1979, Creutz 1980),
$$
d P_{trial}(\phi') \sim \exp\{ - S(\phi')\}  d\phi'~~,
\EQNO{eqheatbath}
$$
i.e. the field $\phi(n)$ is equilibrated in the local heat-bath
provided by its neighbours. In this case the change is always accepted.
The integration over the group, which is needed in order to select
the new field $\phi^\prime$ according to the correct distribution
given by \eq{eqheatbath},
may, however, be difficult to be implemented efficiently
\footnote{*}{It is perhaps worth noting
that the efficiency of an algorithm is not independent of the
architecture of the computer it is implemented on. Up to now, the fastest
computers have been vector machines so that vectorizability is a
prerequisite of a fast algorithm. With the availability of
powerful parallel machines this might change in the future.},
in particular for $SU(N)$ gauge theories with $N \geq 3$.
For those, one has
devised a so-called pseudo-heat bath (Cabibbo and Marinari 1982)
in which
a new $SU(3)$ matrix is obtained from the old one by constructing
$U' = a_2 a_1 U$, where the $a_i$ belong to 2 different
$SU(2)$ subgroups of $SU(3)$ and are subsequently generated
according to \eq{eqheatbath}.

\subsubsection{Critical Slowing Down: Overrelaxed Algorithms}

It has been mentioned already in the previous paragraph that
subsequent configurations in the Markov chain
are not necessarily independent of each other.
The number of steps in the Markov chain over which such a
correlation extends is quantified by the autocorrelation
time $\tau_{\exp}$,
$$
C_{\cal O}(\tau) = \langle {\cal O}(t+\tau) {\cal O}(t) \rangle
\sim e^{-\tau / \tau_{\exp}}. \EQNO{fargo}
$$
In principle, the autocorrelation time depends on the observable
${\cal O}$.
However, with $\tau_{\exp}$, usually the relaxation time of the
slowest mode is meant. One can also define an integrated
autocorrelation time
$\tau_{\rm int,{\cal O}} = {{1}\over{2}} \int d \tau C_{\cal O}(\tau)
/ C_{\cal O}(0)$,
which needs not to be the same as $\tau_{\exp}$ and typically
turns out to be somewhat
smaller than $\tau_{\exp}$. If the number of generated configurations
is $M$ than only $M/\tau_{\rm int,{\cal O}}$
statistically independent measurements of ${\cal O}$
can be performed. Close to a critical point,
i.e. when the continuum limit is approached and a (dimensionless)
correlation length $\xi_L$ diverges,
the autocorrelation time also diverges. Both are related through a
scaling law
$$
\tau_{\exp} \; \sim \; \xi^z_L ~~,
\EQNO{eqcsd}
$$
where $z$ denotes a dynamic critical exponent.
The divergence
of $\tau_{\exp}$ near the continuum limit is called critical slowing
down. Physically a correlation length $\xi_L$ means that a
modification of e.g. a spin at site $n$ influences the orientation of
spins being a distance $\xi_L$ apart. A given numerical update scheme,
however, requires of the order of $\tau_{\exp}$ updates to
communicate changes made at site $n$ over a distance $\xi_L$. Thus
only after $\tau_{\exp}$ updates does one obtain a configuration which
contains new information, \ie~may be considered as independent.
Local algorithms such as
the standard Metropolis as well as the heat-bath
updating proceed through configuration space
like random walks. Correspondingly, their critical exponent
has been determined as $z \simeq 2$. An ideal algorithm, on the other
hand, would lead to $z = 0$. Much research is
therefore going into developing improved methods to reduce
this critical slowing down.

A by now widely used algorithm for non-abelian gauge theories
is able to reduce the critical exponent
to $z=1$. It is based on the transfer of overrelaxation techniques, as
used in solving linear equations, to the Markov process (Adler 1981).
Consider (Brown and Woch 1987, Creutz 1987) a
single link $U$ in $SU(2)$ gauge theory.
The part of the action which depends on $U$ is given by
$$
S = - tr\{ U V \} + {\rm terms}\;\,{\rm independent}\;\,{\rm of} \;\,U~~,
\EQNO{eqmin}
$$
where $V$ denotes the sum of ``staples" around the specified $U$.
Let $U_0$ be the value of $U$ which minimizes the action \eq{eqmin}.
Now construct a new link $U'$ by
$$
U' = U_0 U^{-1} U_0~~.
\EQNO{eqOR}
$$
Then $P_{trial} (U \rightarrow U') = P_{trial} (U' \rightarrow U)$,
so that detailed balance is satisfied if the new link is accepted with
the usual probability, \eq{eqaccept}.
For $SU(2)$, the sum of staples is proportional to an $SU(2)$ matrix
again, and $U_0$, as obtained from $U_0 = {1\over {\det V}} V^{-1}$,
indeed minimizes $S$,
\eq{eqmin}. Moreover, $tr\{U' U_0^{-1}\} = tr\{ U U_0^{-1}\}$
so that the action is preserved under the change and the new link is
accepted with probability $1$. Rewriting \eq{eqOR} in the form
$$
e^{i \vec{u}' \vec{\sigma}} =
e^{i \vec{u}_0 \vec{\sigma}}
e^{-i \vec{u} \vec{\sigma}}
e^{i \vec{u}_0 \vec{\sigma}}
\EQNO{unibi}
$$
shows that the new link lies ``on the opposite side" of $U$ with respect
to $U_0$, i.e. at the maximal distance from $U$. Compared to linear
equation solvers this amounts to choosing the overrelaxation parameter
equal to $2$.
As the given prescription is action conserving,
one has to add standard Metropolis or
heat-bath updates to assure ergodicity.
For $SU(3)$, the sums of staples are not multiples of
an $SU(3)$ matrix and the action is not preserved. One has to perform
the accept/reject decision, \eq{eqaccept}, and also provide some
method to find the minimum of the action, $U_0$.

Loosely speaking, the reduction of $z$ to $z \simeq 1$ is gained by
the right balance between deterministic
and ergodic steps. Further improvements seem to require
knowledge about the long-distance, slowly moving modes
in order to update the corresponding local fields collectively.
For Ising systems and other models with global symmetry,
where an embedding of Ising
spins into the original theory is possible, so-called
cluster algorithms have been devised, which are able to obtain
$z \simeq 0$ (Swendsen and Wang 1987, Wolff 1989, Hasenbusch 1990).
For gauge theories, the local gauge invariance poses a
considerable complication. Moreover, it was shown
that for theories with field variables which take values in other
manifolds than spheres an embedding of Ising-like systems
is not possible (Caracciolo \etal~1992). One therefore has
to search for other
collective-mode algorithms.

\subsubsection{Fermion Simulations: Hybrid Monte Carlo Algorithm}

The existence of fermions in nature presents an additional
complication in numerical lattice calculations. Fermions
follow the Pauli exclusion principle and obey anti-commutation
relations. These would be comprehensible to a computer only
through matrix representations, an approach which is
prohibitively memory consuming. As the fermion part of
the path integral is Gaussian,
$$
S_F = {\overline \psi} (D + m) \psi ,
\EQNO{physik}
$$
with $D$ being the Dirac matrix and $m$ the fermion mass,
one can integrate over the fermion fields analytically,
$$
\int {\cal D} \psi {\cal D} {\overline \psi}
\exp( - {\overline \psi} (D + m) \psi )
= \det \{ D + m \}~~.
\EQNO{theorie}
$$
This manipulation leads to the famous fermion determinant, whose
calculation except for small lattices
is practically impossible because
the number of operations required grows like
(volume)$^3$ and
suffers from numerical instabilities.
However, there are ways to deal with it
in form of trading the determinant for the inverse of the
Dirac matrix. This will be explained in some detail below.
Even then, the numerical effort is quite large so that
many lattice investigations have been using the so-called
quenched approximation. This approximation amounts to setting
the determinant equal to 1. Seemingly a crude
approximation, a quick calculation reveals that
$det \{ D + m \} {\buildrel !\over =} 1$
amounts to neglecting virtual fermion loops,
$$
{\rm tr} \log \{ D+m \} \sim \sum_{k=0}^{\infty} ({1\over m})^{2k} D^{2k}
{\buildrel !\over =}  0 .
\EQNO{fehler}
$$
The Dirac matrix $D$ connects neighbouring lattice sites via
a gauge link and contributes only with even powers
to the expansion, \eq{fehler}, because
otherwise the integral over
the gauge fields vanishes. For the same reason, only closed loops
do not average to zero. Thus, the quenched approximation
neglects (virtual) quark loops and
treats fermions as static degrees of freedom.
Properties
of the theory which depend crucially on the fermion dynamics
are thus not accessible by studies in the quenched approximation.
On the other hand, basic properties of e.g. QCD which are dominated
by the non-abelian gluon dynamics should and do survive the
approximation. Thus, quenched studies may serve as important guides
for many non-perturbative aspects of the theory. Of course, the
results have to be checked by calculations in the full theory.

All fermion algorithms in use, which try to take into account
the dynamical effects of fermions in the generation of field
configurations, re-express the determinant by
a path integral over pseudofermion fields i.e. bosonic fields
which interact via the fermion matrix
(Petcher and Weingarten 1981, see also Fucito \etal 1981).
As the fermion matrix
is not positive definite, one first has to square the determinant
in order to obtain a regular Boltzmann weight factor,
$$
\eqalign{
{\det}^2 \{D+m\} &= \det \{ (D+m)^{\dag} (D+m) \} \cr
                 &=
\int {\cal D} \phi {\cal D} \phi^*
\exp ( \phi^* [(D+m)^{\dag}(D+m)]^{-1} \phi )~~.  \cr}~~
\EQNO{formel}
$$
Thus one has traded the determinant for the inverse matrix at the
cost of simulating twice the number of fermion species.
However, in terms of pseudofermionic variables, the fermions
can now be dealt with in an ordinary Monte Carlo. Because of
the non-locality of the inverse Dirac matrix, any
local updating scheme for the gauge fields would though require
to recalculate the inverse after each local change in the $U$'s.
Alternatively, one could change a whole gauge field configuration
at once and then recalculate the inverse. However, with ordinary,
local updating procedures, the acceptance probability
of a global change would drop to zero very quickly with the lattice
size.

The hybrid Monte Carlo algorithm (Duane \etal 1987), which to date
is the only algorithm
which is exact and also not prohibitively
time consuming (at least on lattices which are not enormously large),
solves this problem by deliberately preparing a new configuration
for a global
accept/reject decision.
This preparation is based on elements taken from previously developed
approximate algorithms.
Suppose one adds a quadratic term to the action,
$$\eqalign{
{\cal H} =& {1\over 2} \sum_{n,\mu} tr \pi_{\mu}^2 (n) + S_E \cr
=& {1\over 2} \sum_{n,\mu} tr \pi_{\mu}^2 (n)
+ S_G(U) + \phi^* [(D+m)^{\dag} (D+m)]^{-1} \phi ~~. \cr}~~
\EQNO{eqhamilton}
$$
The extra term can be integrated out analytically and does
not change vacuum expectation values.
This expression, \eq{eqhamilton}, is now taken as a Hamiltonian
from which equations of motion in a fictitious time $\tau$
are derived.
The $\pi$ fields can be considered as conjugate momenta
to the fields $U$,
slightly modified
to preserve the gauge fields as elements of the gauge group,
$$
\dot{U}_{\mu} (n) = i \pi_{\mu}(n) U_{\mu}(n) .
\EQNO{eqmdUevl}
$$
With the pseudofermion fields $\phi$ kept constant during the
evolution in the time $\tau$
(Gottlieb \etal 1987a), the second set of Hamilton equations is given by
$$
i \dot{\pi}_{\mu}(n) =
\left[ U_{\mu}(n) {\partial\over {\partial U_{\mu}(n)}}
\{ S_G(U) + \phi^* [(D+m)^{\dag} (D+m)]^{-1} \phi \} \right]_{TA} ~~,
\EQNO{eqmdpievl}
$$
where the subscript $TA$ denotes the traceless antihermitian
part in colour space, $A_{TA} = (A-A^{\dag}) - {{1}\over{N}}
tr(A-A^{\dag})$.
This step is, of course, the time consuming one as it requires to calculate
the inverse of the fermion matrix.
With momenta conjugate to the pseudofermion fields added, a pure
Hamiltonian evolution
has been used in the past for simulations of the
full theory (Callaway and Rahman 1982),
exploiting the fact that for large systems the canonical distribution,
$\exp(-{\cal H})$ is very sharply peaked around the
microcanonical one, $\delta(E-{\cal H})$. The advantage
of the Hamiltonian evolution is its relatively fast progressing
through phase space. However, there are reasons to believe
that it violates ergodicity. Alternatively,
Langevin methods (Parisi and Wu 1981, Fukugita and Ukawa 1985,
Batrouni \etal 1985)
have been used, which replace the two first order
differential equations given above, i.e. a second order derivative in
the fields, by a single first order one.
Generically, the Langevin evolution is given by
$$
\dot{\phi} = - {\partial\over {\partial \phi}} S + \eta ,
\EQNO{eqlevl}
$$
which also includes a noise term $\eta$ to
guarantee ergodicity. However, a pure Langevin
evolution is basically a random walk
through phase space and thus progress is rather slow.

In the hybrid Monte Carlo algorithm the system is guided
through phase space by combining
Langevin steps with molecular dynamics iterations.
Thereby the advantages of both, Langevin and microcanonical algorithm,
fast progress through phase space and ergodicity, are utilized.
To be specific, at the beginning of a so-called trajectory,
the momenta as well as the pseudofermion fields are refreshed
in random manner, for QCD (Gottlieb \etal 1987a),
$$
\eqalign{
\pi_{\mu}(n) &= \sum_{a=1}^8 \lambda_a r_{\mu}^a(n)~~, \cr
\phi^i(n) &= \sum_m \sum_{j=1}^3 (D+m)^{\dag}_{ij} (n;m) r^j (m)~~.
\cr }
\EQNO{faculty}
$$
Here, the $r$'s represent (two different) sets of gaussian random numbers
with variance $\langle r^2 \rangle = {1\over 2}$,
the $\lambda$'s are the generators of the gauge group and
$a$ and $i,j$ are colour indices.
Note in passing that this gives the desired distributions of the
$\pi$ and $\phi$ fields. This step is then followed by
a number $N_{MD}$ of molecular dynamics iterations, \eq{eqmdUevl}
and \eq{eqmdpievl}.
At the end of such a trajectory one has
deliberately prepared a configuration which then is
accepted as new configuration with probability
$$
P_{accept} = \min \{ 1, \exp(-[ {\cal H}_{final} - {\cal H}_{initial}])\}~~.
\EQNO{bielefeld}
$$
If the integration of the equations of motions
could be carried out exactly, the acceptance probability
for a configuration at the end of a trajectory would be unity
because energy is preserved in Hamiltonian dynamics.
However, the necessity to solve the equations of
motion numerically, i.e. the introduction of finite time
step sizes $d \tau$, needed to discretize the
$\tau$ derivatives in \eq{eqmdUevl}-\eq{eqmdpievl},
unavoidably introduces integration errors which
lead to slight violations of energy conservation.
These errors can be studied by means
of the Fokker-Planck equation (Batrouni \etal 1985).
It turns out that they can
partially be absorbed into a renormalization of the coupling constants.
But there are also non-integrable terms.
The hybrid Monte Carlo compensates for these errors by
performing the above global accept/reject decision about the accumulated
changes of the configuration at the end of a
trajectory (Scalettar \etal 1986, Duane \etal 1987). This
Monte Carlo element makes the algorithm an exact one. The
evolution along a trajectory is taken only as a way to
very carefully prepare a new tentative update of the system
as a whole.

In order to be a viable Monte Carlo procedure, the hybrid Monte Carlo
algorithm has to maintain detailed balance. This is achieved by
letting evolve the fields and their conjugate momenta according to a
leapfrog discretization scheme.
In its simplest version this is given by
$$
\eqalign{
\pi(\tau + d \tau) &= \pi(\tau) + \dot{\pi}(\tau + {1\over 2} d \tau)
d \tau +  O(d \tau^3) ~~,\cr
U(\tau + {1\over 2} d \tau) &= e^{i \pi(\tau) d \tau}
U(\tau - {1\over 2}d \tau) + O(d \tau^3) ~~. \cr}
\EQNO{eqleapfrog}
$$
As indicated, the errors are proportional to $d \tau^3$ while in the
initial and final half steps of the $U$ field evolution the
error is $O(d \tau^2)$. In the end, this amounts to an
energy violation $\sim V d \tau^4$, where $V$ is the 4-dimensional
lattice volume
(Creutz 1988, Gupta R \etal 1988b, Horsley \etal 1989, Gupta S \etal 1990).
Thus the time step size has to be scaled as $V^{-5/4}$ in order
to maintain a constant acceptance probability.
Of course, this error can be reduced by decreasing
the time step size appropriately but then progress through
phase may become slow. Therefore, it is worthwhile
to think about refinements of the
leapfrog scheme, \eq{eqleapfrog},
(Campostrini and Rossi 1990,
Sexton and Weingarten 1992).
Another important issue concerns the critical behaviour of
the hybrid Monte Carlo. Studies of simpler models like
the 2-D XY model suggest a dynamic critical exponent of
$z \simeq 1$, if the trajectory length is varied
proportionally to the correlation length,
$\tau = N_{MD} \,\delta \tau \sim \xi_L$,
and $z \simeq 2$ for fixed $\tau$ (Kennedy and Pendleton 1990,
Gupta S 1992a).

In simulations with dynamical fermions,
by far the largest fraction of computing time
goes into repeatedly calculating the inverse of the Dirac
matrix. For that purpose, iterative inversion solvers
like the conjugate gradient for staggered and the minimal
residual algorithm for Wilson fermions are used. These schemes
already exploit the sparse nature of the Dirac matrix rather
efficiently. Still,
the number of iterations necessary to reach a certain
accuracy depends on the fermion mass via the condition number
$$
N_{iter} \sim {{|\lambda|_{max}+m}\over{|\lambda|_{min}+ m}}
\sim {1\over m}~~,
\EQNO{chaos}
$$
where $\lambda$ denotes an eigenvalue of the Dirac matrix.
This explains the high cost of simulations with
small fermion masses. Therefore, the investigation of
preconditioning techniques like e.g. ILU decomposition, of
convergence accelerators as e.g. Fourier acceleration or
of multigrid methods is a very active area of research
at present (see for instance Herrmann and Karsch 1991, Kalkreuter 1992).

\subsection{Continuum Limit and Critical Phenomena}
A central question to be answered in any lattice study is
in how far results are influenced by the
lattice discretization or directly represent continuum physics. Clearly
on a coarse lattice with large lattice spacings rotational symmetry is
violated and calculations of, for instance,
correlation functions, will be affected
by this. However, these lattice artefacts should vanish in the continuum limit.

Physical quantities are calculated on the lattice in units of
the lattice spacing $a$; for instance, a mass or inverse correlation
length will be
given in units of $a^{-1}$, the string tension in units of $a^{-2}$ and
the energy density in units of $a^{-4}$. The continuum limit,
$a \rightarrow 0$,
should be taken while keeping physical quantities like the correlation length,
$\xi \sim 1/m_{Hadron}$, fixed. On the lattice the continuum limit
thus corresponds to
a point in parameter space where the lattice correlation
length, $\xi_L = \xi /a$,
diverges. The continuum limit has to be taken at the critical point of the
statistical model defined by the lattice partition function \eq{partlat}.
In fact,
it is a central task of lattice calculations to identify the critical points
in parameter space at which a continuum theory can be defined. Apriori it is
not insured that the QFT defined at such a point coincides with the
perturbatively studied continuum theory. The $SU(2)$-Higgs model as well
as the $U(1)$ gauge theory with fermions on the lattice provide examples
where the known fixed points may lead to trivial non-interacting theories.
We will discuss these models further in section 5. Here we will
concentrate on a discussion of the continuum limit of QCD.

In the case of 4-dimensional QCD with massless quarks,
the lattice action itself does not contain any dimensionful parameter.
The lattice cut-off $a^{-1}$ enters through the renormalization
of the bare couplings, \ie ~the gauge coupling $g^2$
and the quark masses $m_i$.
It is generally expected that the continuum limit of lattice QCD is
reached in the limit $g^2 \rightarrow 0$, where the relation between
$g^2$ and $a$ is known from a perturbative analysis of the renormalization
group equation. To two-loop order one finds for $n_f$ massless fermion flavours
and colour gauge group $SU(N)$,
$$
a\Lambda_L= (b_0g^2)^{-b_1/2b_0^2}~e^{-1/2b_0g^2} ~~,~~\EQNO{rge}
$$
with $b_0,~b_1$ given by
$$\eqalign{
b_0 =&{ 1 \over 16 \pi^2} \biggl[ 11 {N \over 3}- {2 \over 3} n_f
\biggr]\cr
b_1 =&\biggl( { 1 \over 16 \pi^2}\biggr)^2 \biggl[ {34 \over 3} N^2
-\biggl({10 \over 3} N+{N^2-1 \over N}\biggr) n_f \biggr]~~. \cr}
\EQNO{b0b1}
$$
Here $\Lambda_L$ is an unknown scale parameter, not fixed by the theory.
The above relation can then be used to eliminate the
cut-off from calculated, dimensionless quantities ($m_{Hadron}
a,~\sqrt{\sigma}a$)
in favour of the scale parameter
$\Lambda_L$. If the lattice simulations have been performed in the
continuum regime, results should be cut-off independent, \ie~ physical
quantities expressed in units of $\Lambda_L$, like
$m_{Hadron}/\Lambda_L$ or $\sqrt{\sigma}/\Lambda_L$,
should be independent of
the coupling $g^2$ used in the actual calculation.
Numerical studies of the QCD $\beta$-function using Monte Carlo
renormalization group techniques gave evidence that the asymptotic
scaling behaviour, described by \eq{rge}, is approximately valid
for $g^2 < 1$
(Bowler \etal 1986, Gupta R \etal 1988a).
For $g^2 \ge 1$, however, strong violations of
asymptotic scaling have been found. In fact, these violations also gave rise
to the speculation that the continuum limit of lattice QCD will not be
reached at $g^2 = 0$, but may have to be taken at finite $g^2$
(Patrascioiu and Seiler 1992).
The Monte Carlo data we are going to discuss in the following certainly
cannot rule out such a possibility as they can probe only a finite
interval
of $g^2$ values. However, they also do not show any evidence for such
an unconventional behaviour.

Detailed studies of the scaling and asymptotic scaling behaviour have been
performed recently in pure $SU(N)$ gauge theories. The simplest physical
quantities accessible in this case are the string tension, $\sigma$, as
well as the deconfinement transition temperature, $T_c$. Both quantities
have been studied on large lattices over a relatively wide range of couplings,
which, nonetheless, does not allow to go to very small values
of $g^2$.
It turns out that the ratio of these observables, $T_c/\sqrt{\sigma}$,
indeed to a large extent is insensitive to variations of the cut-off.
This is shown in \fig{ratio} for the $SU(2)$ and $SU(3)$ gauge theory.

Averaging over the ratios measured for small values of $aT_c$ one finds
$$
{T_c \over \sqrt{\sigma}}= \cases{
0.69 \pm 0.02 &, SU(2) \cr
0.56 \pm 0.03 &, SU(3) \cr}
{}~.~~\EQNO{ratiots}
$$
It is rather remarkable that scaling of $T_c / \sqrt{\sigma}$ is
valid in the case of $SU(2)$ to a high degree over the entire range of
lattice spacings from $aT_c \simeq 0.25$ on downwards. Using as an
input the experimental value for the string tension,
$\sqrt{\sigma}=(420 \pm 20)$~MeV, this corresponds
to lattice spacings $a \lsim 0.5$~fm,
i.e. only little less than typical hadronic
scales of 1~fm.
For $SU(3)$ the same seems to hold for $a \lsim 0.25$~fm.
A similar behaviour is found in the ratio $T_c / m_{\rho}$,
shown in \fig{tcrho}. Here we have also included results from
simulations with two light quark flavours
($n_f=2$) (Gottlieb \etal 1987b, Bernard \etal 1992d, Gottlieb \etal 1992).
Results for various ratios of physical observables
are summarized in table~\table{ratios}.

\midinsert
\centerline {\tenbf Table~\table{ratios}:}
{\baselineskip=12truept\tenrm The ratio of various physical observables.
The hadron masses have been taken from
(Billoire {\sl et al} 1985, Langhammer 1986) ($SU(2)$, $n_f=0$),
(Kogut {\sl et al} 1991) ($SU(2)$, $n_f=4$),
(Cabasino {\sl et al} 1991, Gupta R {\sl et al.} 1991a)
({\its SU}(3), {\its n}$_f$=0) and (Altmeyer {\sl et al} 1993)
({\its SU}(3), {\its n}$_f$=4). Results for {\its n}$_f$=2 have been
taken from (Gottlieb {\sl et al} 1987b).
With $m_G$ we denote the mass of the
lowest lying glueball state, $0^{++}$.
}
$$
\vbox{\offinterlineskip
\halign{
\strut\vrule     \hfil $#$ \hfil  &
      \vrule # & \hfil $#$ \hfil  &
      \vrule # & \hfil $#$ \hfil  &
      \vrule # & \hfil $#$ \hfil  &
      \vrule # & \hfil $#$ \hfil  &
      \vrule # & \hfil $#$ \hfil
      \vrule \cr
\noalign{\hrule}
&&~T_c/\sqrt{\sigma}~
&&~T_c/ m_{\rho}~
&&~T_c/ m_N~
&&~\sqrt{\sigma} / m_{\rho}~
&&~m_{\rm G} / m_{\rho}~\cr
\noalign{\hrule}
 ~SU(2) ~&&~~&&~~&&~~&&~~&&~~\cr
 ~n_f=0 ~&&~0.69(2)~&&~0.25(3)~&&~-~&&~0.36(3)~&&1.4(3)\cr
 ~n_f=4 ~&&~0.38(2)~&&~0.15(3)~&&~-~&&~0.39(3)~&&1.1(2)\cr
\noalign{\hrule}
 ~SU(3) ~&&~~&&~~&&~~&&~~&&~~\cr
 ~n_f=0 ~&&~0.57(3)~&&~0.31(3)~&&~0.24(3)~&&~0.53(9)~&&1.8(3)\cr
 ~n_f=2 ~&&~-~&&~0.19(1)~&&~0.12(1)~&&~-~&&-\cr
 ~n_f=4 ~&&~0.38(5)~&&~0.17(1)~&&~0.11(1)~&&~0.45(6)~&&-\cr
\noalign{\hrule}}}
$$
\bigskip
\endinsert

Despite of this impressive scaling behaviour,
the lattice simulations clearly have
not been performed in the asymptotic scaling regime defined as the
regime of validity of \eq{rge}.
This becomes obvious
from the behaviour of $T_c/\Lambda_{\rm \overline{\rm MS}}$ shown in
\fig{tc}\footnote{$^{\dagger}$}{We use the more familiar continuum scale
parameter
$\Lambda_{\overline{\rm MS}}$ rather than the lattice scale parameter
$\Lambda_L$. Both are related through
$\Lambda_L / \Lambda_{\overline{\rm MS}}^{N,n_f} = \exp{\{\frac{1}{2b_0}
\left( {\frac{1}{8N}-0.169956N+P_4 n_f } \right) \} }$,
where $P_4 = 0.0026248$ for staggered fermions and
$P_4 =0.0068870$ for Wilson~fermions (Weisz 1981,
Sharatchandra \etal 1981).  }.

The scaling behaviour found for various ratios of physical observables
suggests, however, that the violations of asymptotic scaling found in
individual quantities are of common origin and could be removed in a
suitable renormalization scheme. There have been various suggestions
for the choice of renormalized couplings, which could improve the
asymptotic scaling behaviour (Parisi 1980). In \fig{tc} we also show
$T_c/\Lambda_{\overline{\rm MS}}^{N,0}$ obtained from \eq{rge}
after replacing the bare coupling $\beta$ by
$$
\beta_E = {N^2-1 \over 4 \langle P   \rangle}~~,~~\EQNO{beff}
$$
where $\langle P  \rangle$
denotes the expectation value of the Euclidean action density
given by the plaquette term defined in \eq{plaq}.

The rather weak dependence of physical quantities on the coupling in the
renormalized, $\beta_E$-scheme allows to use physical quantities, known
from experiment, as input and determine the unknown QCD scale parameter.
 From \fig{tc} we deduce for the $SU(3)$ gauge theory
$\Lambda_{\overline{\rm MS}}^{3,0}=( 240 \pm 30)$~MeV
using the string tension value $\sqrt{\sigma} = (420 \pm 20)$~MeV as
input. Similar results have been obtained from studies of the short
distance part of the heavy quark potential (Bali and Schilling 1993,
Booth \etal 1992) and the 1P - 1S fine splitting in the
charmonium spectrum
(El-Khadra \etal
1992), which led to consistent values for
$\Lambda_{\overline{\rm MS}}^{3,0}$. Also in
the case of $SU(2)$ gauge theory the values obtained for
$\Lambda_{\overline{\rm MS}}^{2,0}$ are consistent with the above value
(Fingberg \etal 1992, Michael 1992, L\"uscher \etal 1993).
We note, however, that the physics of the $SU(2)$ gauge
theory clearly is different from that of the $SU(3)$ gauge theory. This
is particularly evident from the ratio $\sqrt{\sigma}/m_{\rho}$, which
can be obtained from Table 1,
$$
{\sqrt{\sigma} \over m_{\rho}} = \cases{
0.36 \pm 0.04 &, SU(2) \cr
0.54 \pm 0.05 &, SU(3) \cr}
{}~.~~\EQNO{ratiosm}
$$
This should be compared with the experimental value, $\sqrt\sigma /
m_\rho = 0.545\pm 0.026$.
Thus quenched QCD (pure $SU(3)$ gauge theory)
provides a surprisingly good approximation
to the real world in this case. On the other hand, a strong
dependence of $\Lambda {{3, n_f}\over{MS}}$ on the number of flavours
is known from perturbative analysis of high energy data.
The calculation of the string tension for four flavour QCD (Born \etal
1991a) provides indications from the lattice that dynamical quarks
reduce the value of $\Lambda {{3, n_f}\over{MS}}$ substantially,
leading to $\Lambda {{3 , 4}\over{MS}} = (150 \pm 40)$~MeV in the
$\beta_E$ scheme.

\section{QCD at Low Energies}

Quantum Chromo Dynamics is generally accepted as the microscopic
theory of the strong interactions. At large momentum transfers, due
to asymptotic freedom, QCD has been exploited perturbatively with
great success. At low energies, however, the effective
quark-gluon coupling becomes large and non-perturbative
techniques have to be adopted. Numerical lattice investigations
provide a unique possibility to study the long-range properties of
QCD with, in principle, no approximation. The aim of lattice QCD is
to collect convincing evidence for anticipated basic features of QCD
at large distances, in particular for confinement and spontaneous
chiral symmetry breaking. Moreover, it is attempted to verify
phenomenological concepts and to predict parameter for
experiment and model building which are inaccessible otherwise.
Naturally, one has to calibrate the precision of lattice investigations
by confronting lattice results with experimental data.

\subsection{Confinement and the Heavy Quark Potential}

Charmonium and in particular bottonium systems are
sufficiently heavy to allow non-relativistic concepts
to be adopted. The interactions between these heavy quarks
can then be described by means of a potential.
Also the
confinement hypothesis of QCD can be expressed in a simple
way namely in terms of a potential
which grows when the distance between heavy quarks
is increased.

Phenomenologically, the potential is decomposed
into a spin-independent part which is responsible for
confinement, and fine-structure corrections which
involve spin-orbit and spin-spin interactions.
The spin-independent part $V_C(R)$ of the potential
is empirically rather well described by a superposition
of a term which rises linearly with distance and leads
to the confinement of quarks, plus a Coulomb-like
correction which is motivated by one-gluon exchange
dominating at small distances. The prototype of these
potential ans\"atze is given by the Cornell potential
(Eichten \etal 1975),
$$
V_C(R) = - \frac{4}{3} {\alpha\over R} + \sigma R ,
\EQNO{plus}
$$
with a fixed Coulomb coefficient $\alpha$
and $\sigma$ being the string tension,
$\sqrt{\sigma} \simeq 420$~MeV.
Modifications include a perturbatively running Coulomb
coefficient $\alpha (R)$ at small $R$ and various
interpolation prescriptions between the small and large
distance behaviour (for an overview see e.g. K\"uhn and Zerwas
1988).
In the distance range probed by charmonium and
bottonium states the various ans\"atze are in numerical
agreement. The differences show up in the short distance
properties. This region will be probed by toponium, if
the top quark is not too heavy.
At present, quantities which depend on
wave functions at the origin like e.g. leptonic
widths favor a logarithmically softened Coulomb singularity
as in QCD-like potentials (Buchm\"uller and Cooper 1988).

The potential is obtained in a gauge-invariant way
from the Wilson loop (Wilson 1974),
$$
W(R,T) = \langle e^{ i \oint_{R \times T} d x_{\mu} A_{\mu} } \rangle
\sim e^{-V_C(R) T},
\EQNO{eqwilsonloop}
$$
which describes the propagation of a
static quark-antiquark pair,
separated by a distance $R$,
in Euclidean time $T$.
In \eq{eqwilsonloop} it has been assumed
that $T$ is sufficiently large so that the
ground state potential $V_C$ dominates the expectation value
and excitations of the string of
colour fields between the heavy quarks are exponentially suppressed.

The confinement potential has been the target of quite a few
investigations, starting with the first numerical
lattice studies of Creutz on $5^4$ lattices (Creutz 1980).
As a major technical improvement,
it was possible to
construct operators for the colour string
which have an increased overlap with the ground state
(Teper 1986, Albanese \etal 1987).
This allows to reliably extract
$V_C$ already from Wilson loops with smaller time extent.
The results of recent analyses in the quenched approximation
(Bali and Schilling 1992, Booth \etal 1992)
on lattices up to size $32^4$
are shown in \fig{figqupot}. The data reach distances almost up to
$R = 2$~fm and demonstrate a linear rise at large separations
very nicely. This result as well as observations in earlier attempts
represent important support for the confinement
hypothesis.

The linear rise at large distances is accompanied by a
Coulomb-type behaviour at small $R$. At not too small quark-antiquark
separations, the coefficient of the Coulomb term is compatible
with a universal strength $\pi / 12$ predicted by string models
(L\"uscher \etal 1980) for distances $R \geq 0.3$~fm (Alvarez 1981).
Below $0.1$ to $0.2$~fm it is expected that the
potential is adequately described by
perturbative 1-gluon exchange (Fischler 1977, Billoire 1980),
$$
V(R) = -{{N^2-1}\over{2N}} {{\alpha(R)}\over{R}}  \EQNO{boticelli}
$$
with
$$
\alpha(R) = {{4 \pi}\over{b_0 \ln(1/R^2 \Lambda^2_{\overline {MS}})}}
\times \left[
1- {{b_1}\over{b_0^2}}
{{\ln\ln(1/R^2 \Lambda^2_{\overline {MS}})}\over
{ \ln(1/R^2 \Lambda^2_{\overline {MS}})}}
+ {{(31N-10n_f)/9b_0 + 2\gamma_E}\over
{ \ln(1/R^2 \Lambda^2_{\overline {MS}})}} \right]
\EQNO{eqpertpot}
$$
where $b_0$ and $b_1$ are given in \eq{b0b1}.
Indeed, quantitative analyses of the
short distance behaviour (Michael 1992, Bali and Schilling 1993,
Booth \etal 1992) do find evidence for a ``running" $\alpha(R)$
in accordance with \eq{eqpertpot}.
By setting the scale from the string tension,
the QCD $\Lambda$ parameter can be extracted from the measured
$\alpha (R)$. One finds,
in physical units,
$\Lambda_{\overline{MS}}^{3,0} = (240 \pm 30)$~MeV (for $SU(3)$).
This result, obtained from the short distance properties
of the theory, compares successfully with the results
of other approaches mentioned in section 2.3.
Moreover, a similar analysis in quenched $SU(2)$ is in
agreement with a program
in which
long and short distances are covered in a step-wise
manner, by working on a variety of different
lattices sizes and matching them by means of a variant of the
renormalization group procedure (L\"uscher \etal 1993).
Thus the originally non-perturbative lattice approach is
becoming capable of reaching into the short distance regime
where results can be matched directly to perturbative
calculations.
However, compared to recent LEP data (or to phenomenologically
successful potentials)
the lattice results for the running coupling constant
are too low.
One and presumably the
main reason for this discrepancy is expected to be the
quenched approximation. The coefficient $b_0$ is strongly
dependent on the number of flavours $n_f$, leading to a
difference of about 30 \% between $n_f = 0$ and $n_f = 4$.
This effect will partially be counteracted by an unknown
shift in $\Lambda_{\overline{MS}}^{3,n_f}$.
Moreover, the quenched
string tension from which usually the scale is taken, is not
a physical quantity.
Equating it to the
``experimental" value of 420~MeV neglects the effects of quark
loops. In particular, the potential in full QCD is expected to
flatten out at some distance due to quark pair creation.
This could also affect the numerical
value of the string tension (see, however, \eq{ratiosm}).

Results from simulations of full QCD, \ie~with light
dynamical quarks taken into account, are beginning to emerge
from rather large lattices.
The currently accessible distances are in the
range $0.1 \leq R \leq 1.0$~fm so far.
For those quark-antiquark separations,
investigations in the full theory
(Born \etal 1991a) confirm a Coulomb plus linear form of the potential
and thus support standard heavy quarkonium phenomenology.
At distances larger than 1~fm
one hopes for indications that the flux tube
connecting the heavy quarks breaks. This is expected
because of the spontaneous
creation of light quark-antiquark pairs which screen the
colour charges of the heavy quarks.

Fine and hyperfine splittings in quarkonium spectra, beyond
their phenomenological importance, are interesting because
they depend on the spin-structure of the confining
Bethe-Salpeter kernel. They can be derived from
the spin-dependent potential, the
most general form of which
is obtained from a systematic
expansion in the inverse quark mass $1/m$ and is given
by the Breit-Fermi parameterization,
$$
\eqalign{
V_{spin}(R) & =  {1\over{2 m^2 R}} \;\;\; {\vec L}{\vec S}
                  \; \; \; ( \; V_0'+2V_1'+2V_2'\;) \cr
         & +  {1\over m^2} \;\;\; [ \; {1\over R^2}
             \; ({\vec R}{\vec S_1})({\vec R}{\vec S_2}) -
             {1\over 3} {\vec S_1}{\vec S_2} \; ] \; \; \; V_3 \cr
         & + {2\over{3 m^2}} \;\;\; {\vec S_1}
{\vec S_2} \; \; \; V_4 ,\cr}
\EQNO{weniger}
$$
here written for the case of two identical quark masses $m$.
The three contributions represent subsequently the
spin-orbit, the tensor and the spin-spin parts. The components
$V_i$ can be extracted from correlations among the
colour-electric and colour-magnetic fields (Eichten and Feinberg 1981)
in
Wilson loops, $V_{1,2} \sim \langle B_i E_j \rangle_W$
and $V_{3,4} \sim \langle B_i B_j \rangle_W$.
$V_1$ and $V_2$ are related to the confinement potential
through the Gromes identity (Gromes 1984),
$V_2 - V_1 = V_C$, so that
$V_1$ and $V_2$ can not be of short range both at the same time.
In fact, if the confining Bethe-Salpeter kernel
has a scalar Dirac structure, a long-range component
resides in $V_1 \simeq - V_C$ while for vector confinement
it resides in $V_2 \simeq V_C$. As a consequence, the long-range
part of the spin-orbit potentials is markedly different.
Experimentally, the ratios of the
$P$-level fine splittings in charm and bottom decays allow one to
differentiate between the two cases. They strongly favor
scalar confinement while pure vector confinement seems to be ruled out
by the data (Buchm\"uller and Cooper 1988).

The lattice analyses (Campostrini \etal 1986, Huntley and Michael 1987,
Laermann \etal 1992)
conform with this picture very nicely.
\Fig{figspinpot} summarizes the
findings of a study in full QCD (Laermann \etal 1992).
While $d V_1/ d R$ is flat within
the error bars, $d V_2 / d R$ and $V_3$ drop rapidly and
appear to be well described by perturbative one-gluon exchange.
The spin-spin term is practically zero for
$R \geq \sqrt{2} a$, which is compatible with the perturbative
behaviour $V_4 (R) \sim \delta (R)$.

\subsection{Chiral Symmetry}

The spectrum of QCD and many transition amplitudes of light
quark states reveal the characteristic pattern of a theory whose chiral
symmetry is spontaneously broken. For zero-mass
quarks the action is invariant under separate transformations
of left-handed and right-handed quarks. However,
this global chiral symmetry is not observed in the
hadron spectrum. Instead one finds the typical features
of spontaneous symmetry breaking: a pion whose mass is
nearly zero on a typical hadronic scale, $m_{\pi}^2 / m_{\rho}^2
= 0.04$, which thus might be interpreted as the Goldstone boson of
the broken symmetry, as well as a non-vanishing value for the
quark condensate in the vacuum.
Both features are clearly of non-perturbative nature and invite
a first principle lattice investigation.

The result of a recent investigation of theses questions
in full QCD with staggered fermions (Altmeyer \etal 1993)
is shown in \fig{figchiral}.
At vanishing quark mass the numerical
algorithms to invert the Dirac matrix
$$
\langle {\overline \psi} \psi \rangle =
tr (D + m)^{-1},
\EQNO{mehr}
$$
do not work. One thus
has to calculate at finite quark mass and then to extrapolate
to the chiral limit. From chiral perturbation
theory (Gasser and Leutwyler 1982), it is
expected that the pion mass squared is linear in the quark mass.
Indeed, the lattice data for the pion mass, shown in
\fig{figchiral}, follow a linear quark mass dependence
rather nicely. Moreover, in the chiral limit, at zero quark mass,
the pion mass vanishes.
Furthermore, the quark condensate
has a finite intercept at zero quark mass. The numerical value of
the renormalization group invariant condensate
for one flavour is obtained as
$\langle {\overline u} u \rangle^{RGI} \simeq 200$~MeV with a
10 \% error, quite close to the 190~MeV
extracted by means of
current algebra techniques from low-energy pion reactions.
Moreover, one can calculate
the pion decay constant $f_{\pi}$ from the appropriate
matrix element and exploit the
Gell Mann-Oakes-Renner formula,
$f_{\pi}^2 m_{\pi}^2 = m \langle {\overline u} u
+ {\overline d} d \rangle$,
in the limit of vanishing
quark mass in order to check this value.
Within 10 to 20 \% error agreement has been found.

\subsection{Meson and Baryon Spectroscopy}

A central problem of QCD at low momentum is the computation
of the hadron spectrum. As most of the low-lying hadron masses
are known from experiment very precisely, this branch of lattice
calculations may have limited predictive power as far as purely
numbers are concerned. Nevertheless, these calculations are
important to verify that the
non-perturbative dynamics of QCD at low energies
explains the experimental spectrum and
to control the degree of accuracy reached
in a numerical investigation.

The masses of hadrons are obtained from
investigating the propagation of hadrons in (Euclidean)
time, $\langle H(t) | H^{\dagger}(0) \rangle$. By inserting a
complete set of energy states and projecting to
zero momentum,
one can extract the mass of
the lightest hadron with the right quantum numbers
from the exponential decay at large times
$$
\langle H(t) | H^{\dagger}(0) \rangle \rightarrow e^{- m_H t}.
\EQNO{eqmasses}
$$
Generally, the operators $H$ are sensitive to all hadrons with
a given set of quantum numbers specified by $H$. The matrix element,
\eq{eqmasses}, thus contains contaminations of excited states so that
a large time extent is required in order not to overestimate the mass
of the lightest hadron. Alternatively, one tries to construct operators
with a large projection onto the ground state hadron. These are
typically extended operators which approximate an anticipated S-wave
like shape of the wave function.

A quantity that is very sensitive to effects due to finite lattice spacing,
$a$, and finite spatial volume, $N_\sigma$,
is the ratio of the nucleon
to the $\rho$ mass. This ratio is usually plotted as a function of
the ratio $m_{\pi} / m_{\rho}$, which is a measure of the quark mass.
Recall that one can choose the quark mass freely
in a lattice calculation.
In practice it
is very difficult to perform calculations with
small quark masses. One thus usually has to
extrapolate to the physical quark mass value. The so-called Edinburgh
plot is a way to display whether and how the experimental value
$m_N / m_{\rho} = 1.22$ is approached when the quark mass is decreased
from infinity, where $m_N / m_{\rho} = 3/2$ describes the static limit.
This is shown in \fig{figedin} for Wilson quarks
(Allton \etal 1992a, Butler \etal 1992, Daniel \etal 1992, Guagnelli \etal
1992).
Similar results but with
larger error bars are obtained in the staggered discretization
(Cabasino \etal 1991, Gupta R \etal 1991a, Fukugita \etal 1992a).
\Fig{figedin} contains (part of) the data from lattices with
lattice spacing $a < 0.1$~fm and $N_\sigma a \geq 2$~fm.
While for $N_\sigma a \geq 2$~fm finite volume effects seem to be
under control in this quantity, larger lattice spacings as well as
smaller lattices shift the $m_N / m_{\rho}$ ratio to larger values.
Although, in present lattice calculations,
$m_N / m_{\rho}$ is always found to be larger than the
experimental value,
it is clear that it is decreasing as the quark mass
decreases. Close to the static limit, $m_{\pi} / m_{\rho} \simeq 1$,
the nucleon to $\rho$ mass ratio is significantly above 1.5 in accordance
with potential model calculations. Extrapolations
to the physical pion mass produce remarkably good
results.
For instance, the GF11 collaboration obtains an extrapolated value of
$1.28(7)$ which is consistent with the experimental number within 5~\%.
[The QCDPAX collaboration (Iwasaki \etal 1992a)
obtains somewhat larger $m_N / m_{\rho}$
ratios at small quark masses. They do not work with extended
operators though but rather concentrate on studying contributions
of excited states.]
The close agreement is a bit surprising because
below threshold, i.e.
at pion masses less than $m_{\rho} /2$, the $\rho$
should decay into two pions and a qualitatively
new behaviour should set in.
The $\rho$-meson decay cannot, however, be reproduced
properly in the quenched approximation.
Moreover, in the quenched approximation
pion clouds are different from full QCD which should also
effect the $m_N / m_\rho$ ratio.
Current investigations
with dynamical quarks
(Altmeyer \etal 1993, Bernard \etal 1992c,
Bitar \etal 1990a, 1992, Brown \etal 1991,
Fukugita \etal 1992c, Gupta R \etal 1991b)
do not yet reach sufficiently small
quark masses. Consequently,
little difference to quenched results is seen so far
in the currently
accessible range of $m_{\pi} / m_{\rho}$ values.

As far as other low lying hadrons are concerned, the ordering
of the states seems to be reproduced quite well. The precision
of the numbers, however, is not quite as impressive as for
the nucleon, $\rho$ or $\pi$, although especially the results
for the $\Delta$ resonance have improved considerably in the last
year, $m_{\Delta} / m_{\rho} = 1.63(7)$ (Butler \etal 1992).
Another very interesting candidate is the $\eta'$ because of its
relation to the $U(1)$ problem. Unfortunately, the analysis
of this state requires the calculation of a quark-line
disconnected diagram which is numerically utterly
difficult so that so far no reliable results can be quoted.

Additional work in this area has been concerned about
lattice intrinsic cross checks. In particular,
finite size effects in the presence of dynamical quarks
have been analyzed (Fukugita \etal 1992c, Bernard \etal 1992c).
As the staggered
discretization of quark fields introduces certain
flavour non-diagonal terms at finite lattice spacing
it also is important to verify the restoration of flavour
symmetry
(Altmeyer \etal 1993, Fukugita \etal 1992c).
Likewise, at large lattice spacing the lattice
causes distortions of the continuum dispersion relations.
It has been shown that these effects
are small at current values of the lattice spacing.
Moreover, one has checked quenched Wilson fermion results by
working with a variant of the Wilson action in which
$O(a)$ corrections to the continuum action are reduced
to $O(a^2)$ or $O(g^2 a)$
(Symanzik 1983a, b, Sheikoleslami and Wohlert 1985).
This should not effect the true
continuum limit but it was observed that already at lattice
spacings $a \simeq 0.15$~fm the improved action reproduces mass ratios
which are otherwise only obtained at smaller lattice spacings,
$a \leq 0.1$~fm. At small $a$ little difference was found.
The improved action is, however, somewhat superior in the calculation
of hyper-fine splittings (Allton \etal 1992a, b, El-Khadra 1992,
Lombardo \etal 1992, Martinelli \etal 1992).

\subsection{Glueball Spectroscopy}

A very unique topic for QCD lattice investigations is the spectrum of
bound states purely made of glue, the glueballs.
This is a difficult problem numerically so that
the construction
of suitably shaped operators
(Teper 1986, Albanese \etal 1987)
with large overlaps to the lowest lying glueball states
has been an important step forward technically.
In combination with
variational techniques it is now possible to extract
reliable numbers for glueball masses.
As still a very large number of
independent lattice configurations is needed to obtain
a sufficiently clean signal, these studies are mostly
restricted to the quenched approximation. The numerical results
(see the review by
van Baal and Kronfeld 1989)
could be tied to analytic calculations in small and intermediate
volumes (Koller and van Baal 1988, Vohwinkel 1989)
and very nice agreement was found. In particular,
the apparent disagreement between results obtained from
operators which belong to
two different representations of the lattice rotation group
which are both embedded in the same $2^{++}$ continuum representation
could be resolved this way and traced back to be an effect
present only in small volumes, \fig{figglue}.
It is now established
that the
lightest glueballs are the $0^{++}$ and the $2^{++}$ states,
with $m_{2^{++}}\simeq 1.5 m_{0^{++}}$ while
$m_{0^{++}} \simeq 3.5 \sqrt{\sigma}$
(Michael and Teper 1989).
All other glueball states
are heavier than 2~GeV, in particular the exotic states with
quantum numbers which can not be realized in the quark model.
This is somewhat unfortunate since such a state could not mix with
ordinary ${\overline q} q$ mesons and thus would give a clear experimental
signal. For the non-exotic states large mixing effects can occur
and are presumably strongly enhanced by virtual quark loops.

\subsection{Weak Matrix Elements}

Weak interaction processes involving hadrons receive
long-range QCD corrections due to the binding of
quarks inside hadrons. Ambiguities in various
approaches to these non-perturbative effects
must be resolved in order to settle important
issues as e.g. the $\Delta I = 1/2$ puzzle or the
values of the Cabibbo-Kobayashi-Maskawa (CKM) mixing
angles. Lattice simulations attempt to calculate
the hadronic matrix elements that appear in the
analyses of these questions from first principles.
The required theoretical framework is quite
involved because of complicated renormalizations
and mixings of the operators on the lattice.
Moreover, the coefficients of the renormalized
lattice operators have to be related to the
corresponding coefficients as
calculated in some continuum renormalization scheme.
This matching can be done more reliably with the
perturbatively improved actions mentioned already in
section 3.3 (Martinelli \etal 1991) and improved
(lattice) perturbation theory (Lepage and Mackenzie 1992).
Most of the numerical calculations so far have
been performed in the quenched approximation.
The matrix elements
studied include decay constants,
form factors in semi-leptonic decays, mixing
amplitudes for e.g. $K^0 - {\overline K}^0$
and non-leptonic meson decay amplitudes.
Here we shall confine ourselves to a few examples.

Phenomenologically,
it is quite important to resolve an ambiguity in the
leptonic decay constant $f_B$ of the $B$ meson.
Analyses of $K - {\overline K}$ and $B - {\overline B}$
mixing find two solutions for $\cal CP$ violation in the
CKM matrix which are distinguished by small ($< 150$~MeV)
and large ($> 150$~MeV) values for $f_B$.
The calculation of $B$ meson
amplitudes represents a technical problem
on the lattice because current
lattices with $1/a \leq 3.5$~GeV are rather coarse
for a $b$ quark with mass $m_b \sim 5$~GeV. In order
to circumvent this difficulty one either has worked in
the static approximation where the $b$ quark is held
fixed in space and propagates only in time (Eichten 1988) or
has simulated at various quark masses
around the charm quark mass and extrapolated
to the physical $b$ quark mass
(Gavela \etal 1988a, Bernard \etal 1988).
Some results are
presented in \fig{figfb} where
$\hat{f}_{PS} = f_{PS} \sqrt{m_{PS}}$ is plotted
as function of the inverse mass $1/m_{PS}$ of a
pseudoscalar meson. There is some ongoing discussion
about certain technical details, which fortunately have only a small
effect on the final results for $f_{PS}$, so that
between the different
groups (Gavela \etal~1988a, Allton \etal~1991,
Alexandrou \etal 1991, Abada \etal 1992, Bernard \etal 1992e, Hashimoto
and Saeki 1992)
there seems to be agreement now
that there are
large corrections to the asymptotic scaling law
$f_{PS} \sqrt{m_{PS}} \sim const$, predicted to hold
in the limit of infinitely heavy quarks. Moreover, if
the results stay as they are, lattice QCD predicts
a rather large value for $f_B$, $f_B \simeq 200 \pm 20$~MeV.

Another source of information on the CKM matrix elements are
semi-leptonic decays of $D$ and $B$ mesons.
In order to extract the
mixing angles from experimental data one needs to know
form factors of weak currents between the mesons. So far,
lattice methods have been tested on $K$ and $D$ decays as
a preparation for the phenomenologically more interesting
$B$ case.
The matrix element is computed at several momentum transfers.
Assuming pole dominance for the momentum dependence, the lattice
results for $f_+(0)$
(Lubicz \etal 1991,
Bernard \etal 1991)
are in good agreement with the experimental value which is obtained
by assuming $| V_{cs} | = 0.975$ (Anjos \etal 1989,
Adler \etal 1989).
For $D$ decays into vector mesons, $D \rightarrow K^*$,
the situation is more complicated. For one of the
three relevant
form factors, for
$A_2(0)$, both the lattice investigations as well as the experimental
analyses are separately contradicting each other. (Lubicz \etal 1992)
and E691 (Anjos \etal~1990) obtain
$A_2(0)$ vanishingly small while
(Bernard \etal 1992a)
and E653 (Kodama \etal 1992) report a value definitely different
from zero. On the lattice side, the disagreement lies not so much in
the raw numerical data but in the way the extrapolations are
carried out.
For the other two form factors, $V(0)$ and $A_1(0)$,
the two lattice studies agree more or less
with each other and with experiment.
In particular at the end-point of the leptons'
momentum distribution,
$q_{\max}^2$, both groups
obtain $A_1(q^2_{\max}) \simeq 1$.
It has been suggested (Bernard \etal 1992b) to clarify
the experimental situation especially at the end-point
so that the unambiguous lattice prediction can be checked
for systematics before the $A_2$ discrepancy is further analyzed.

The Kaon $B$ parameter is defined by
$$
B_K =
\langle {\overline K} | {\cal O}^{\Delta S = 2} | K \rangle /
{8\over 3} f_K^2 m_K^2~~,
\EQNO{eqkaonb}
$$
where the denominator is the vacuum saturation value.
$B_K$ is related to the ${\cal CP}$ violation parameter
$\epsilon$ in the $K^0$ system and a determination of $B_K$
allows constraints on the mixing angles and the top mass.
The Kaon $B$ parameter has been calculated both for the
Wilson
(Gavela \etal 1988b, Bernard and Soni 1990, Gupta R \etal 1992)
and the staggered (Sharpe \etal 1992b)
discretization of the quark action.
Rather accurate results could be obtained in the latter scheme because
a continuous chiral symmetry on the lattice protects
the operator from mixing with operators of
left-right chiral structure.
These mixings do occur for the Wilson case and have to be
subtracted. The results for the two discretization schemes show
reasonable agreement. There are sizable $O(a)$
corrections which have been studied during the last two years,
the result being that $B_K \simeq 0.6$
in the limit $a \rightarrow 0$.  The analyses have been
repeated in full QCD now (Kilcup 1991, Fukugita \etal 1992b),
with little difference to the quenched
results being seen.

Finally, a number of attempts have been made to compute the
$K \rightarrow \pi \pi$ amplitudes in order to
understand the $\Delta I = 1/2$ rule and to constrain
the ${\cal CP}$ violating angle in the CKM matrix
through the parameter $\epsilon'$. Despite some original optimism,
these quantities have been proven hard
to be extracted from the lattice. The $\Delta S = 1$ four
quark operators mix with a number of operators with lower
dimension. Their contributions have to be subtracted
non-perturbatively which has turned out to be
difficult numerically. For staggered quarks with
their chiral properties the situation is somewhat better.
Here one can use chiral perturbation theory to relate
the $K \rightarrow \pi \pi$ amplitudes to
those for $K \rightarrow \pi$
and $K \rightarrow vacuum$ and fix the subtractions
by chiral symmetry.
Still, the contributions
of so-called eye-operators are not stable numerically.
The main problem, however, is that whether one computes
$K \rightarrow \pi \pi$ or $K \rightarrow \pi$,
one has to deal with final state $\pi \pi$ interactions.
This is difficult. It has
been shown (Maiani and Testa 1990) that one
can not just put two pions with
non-zero momentum on a Euclidean lattice
and expect them to represent an ``out" state.
So one is beginning to first investigate
$\pi \pi$ systems (Guagnelli \etal 1990, Sharpe \etal 1992a)
before one can come back to the original problem.

\section{QCD at High Temperature} \ssf
\subsection{A New Phase of Matter}

One of the outstanding predictions of QCD, which still awaits its experimental
verification, is the existence of a new phase of strongly interacting matter
at high temperatures and/or densities -- the Quark Gluon Plasma (QGP):
At temperatures of the order of the pion mass, $T \sim 140 \MEV$, or
densities a few times that of ordinary nuclear matter,
$n \sim (3-5) n_0$, with  $n_0= 0.15{\rm fm}^{-3}$, conditions are reached
where hadrons start overlapping and the interaction among the
internal partonic degrees of freedom, described by QCD,
becomes important. Theoretical investigations of this complicated
transition regime, in particular the phase transition itself, clearly
require non-perturbative techniques. However, somewhat surprisingly, it
turned out that such an approach is important also in the high
temperature phase. Despite asymptotic freedom a perturbative
description of the plasma phase itself is not at all obvious;
the perturbative expansion is plagued by infra-red divergences
(Polyakov 1979, Linde 1980 and Kapusta 1979),
which become worse with increasing
order of the expansion. In particular, one finds that in the gluon
propagator an electric screening mass of $O(gT)$ and a magnetic
screening mass of $O(g^2T)$ have to be generated dynamically in order to
render the expansion of thermodynamic quantities
finite. These mass gaps signal the existence of
different length scales in the plasma phase.
In addition to the characteristic perturbative length scale, $l_P \sim
1/T$, a hierarchy of non-perturbative length scales like the electric
($l_E$) and magnetic ($l_M$) mass gap show up,
$$\eqalign{
l_P  & < ~~l_E~~   < ~~l_M ~~  \cr
1/T  & <  1/gT     < 1/g^2T } ~~,~~\EQNO{qgpscales}
$$
and thus limit the applicability of perturbation theory.
A perturbative treatment of long-distance properties of the QCD
plasma phase seems to be impossible. In fact, the above relations led to doubts
about any perturbative treatment of the high temperature phase in terms of
quarks and gluons as basic excitations, and it has been suggested that a
description in terms of colourless quasi-particles may be more
appropriate (DeTar 1988).

Different physical observables are sensitive to different momentum regimes.
They thus
may be influenced differently by the various length scales. This, for instance,
shows up already in the equation of state (EOS). While the energy density
rapidly approaches ideal gas behaviour, the pressure shows very strong
deviations up to temperatures $T \sim (2-3)T_c$. Some results for the EOS
of a pure $SU(3)$ gauge theory (Engels \etal~1990a) and two flavour QCD
(Gottlieb \etal~1987d) are shown in \fig{eossu3}.
We note that although the phase transition is first order in the case
of quenched QCD (pure gauge theory)
also the energy density decreases strongly close to $T_c$. This leads to
a rather small latent heat of the transition. Measured in units of the
energy density of an ideal gluon gas, $\epsilon_{\rm SB}$, one finds
(Engels \etal~1990a, Iwasaki \etal~1991, 1992b)
$$
{\Delta \epsilon \over \epsilon_{\rm SB}} = 0.315 \pm 0.030~~.\EQNO{latent}
$$
Using, for the critical temperature of the deconfinement
transition in the $SU(3)$ gauge theory, a value of $T_c \simeq 235$~MeV (from
Table 1) yields
$\Delta \epsilon \simeq 635$~MeV/fm$^3$
for the latent heat.

\subsection{Deconfinement}

Physically it is rather tempting to relate the finite temperature QCD
phase transition to the deconfinement mechanism. Quarks and gluons are
confined at low temperature and form colourless hadrons. These may break up
at high temperatures and the partons become free, i.e. deconfined. This
picture can be made rigorous in pure $SU(N)$ gauge theories, where
dynamical quarks are absent.
Dynamical quarks will spoil the strict notion
of confinement even at zero temperature
as quark-antiquark pair creation will lead to a flat heavy quark potential
at large distances. The potential thus would not be confining in the strict
sense.

Pure $SU(N)$ gauge theories have a global $Z(N)$ symmetry, which controls
the properties of the finite temperature transition.
If one performs a global $Z(N)$ rotation of
all timelike gauge fields $U_0(n)$ originating from the sites
$n=(n_0,\vec n)$ of a given temporal hyperplane (fixed $n_0$)
of the lattice,
$$
U_0(n_0,\vec n) \rightarrow U'_0(n_0,\vec n)=
zU_0(n_0,\vec n)~~,~~z\in Z(N), ~n_0~{\rm fixed}~~,~~\EQNO{tlink}
$$
the action remains unchanged, $S_G(\{U'_{\mu}(n) \})=
S_G(\{U_{\mu}(n)\})$. On the contrary, the Polyakov loop,
$$
L_{\vec n}= \prod_{n_0=1}^{N_{\tau}}
U_0(n_0,\vec n)~~,~~\EQNO{poly}
$$
which describes the propagation of
a static fermionic test charge and probes
the screening properties of the surrounding gluonic medium,
transforms non-trivially under this transformation,
$$
L({\vec n}) \rightarrow zL({\vec n})~~,~~z\in Z(N)~~.~~\EQNO{trans}
$$
Its expectation value, $\langle L({\vec n}) \rangle$,
thus will vanish as long as the theory preserves the global $Z(N)$
symmetry.
It will, however, acquire a non-vanishing value if this
symmetry is spontaneously broken.

It is expected that the critical properties of $(d+1)$-dimensional
gauge theories are described by an effective $d$-dimensional spin model
given in terms of the Polyakov loop. Critical indices should then only
be determined by the dimensionality of the spin system and its global
$Z(N)$ symmetry (Svetitsky and Yaffe 1982a, b).
In fact, there is strong
evidence from numerical simulations that the phase transition for $SU(2)$ gauge
theory is of second order and in the same universality class as
the 3-$d$ Ising model, whereas the $SU(3)$ gauge theory leads to a
first order transition, as expected on the basis of these general
universality arguments.

The critical properties at the transition point have been studied in detailed
finite size scaling studies, which allow a rather accurate determination of
critical indices. In \fig{decon} we show the result of such a scaling analysis
for the critical exponent $\nu$ of $SU(2)$
(Engels \etal~1990b) and $SU(3)$ (Fukugita \etal~1990)
gauge theories. In particular one finds:
$$\eqalign{
SU(2):&\quad {\beta \over \nu}~=~0.545\pm0.030\quad
{\gamma \over \nu}~=~1.93\pm0.03\quad \nu~=~0.65\pm0.04\quad  \cr
SU(3):&\quad {\gamma \over \nu}~=~3.02\pm0.14\quad\quad
\nu~=~0.34\pm0.01\quad  \cr}
$$
While the exponents for $SU(2)$ are in agreement with those of the
three dimensional Ising model, $\beta / \nu =0.516(5)~,~~ \gamma / \nu=
1.965(5)$, $\nu=0.63(3)$,
the result for $SU(3)$ is consistent with $\nu = 1/3$, $\gamma = 1.0$,
which is expected for a first order phase transition in 3-dimensions.

Further inside into the nature of the phase transition in the pure gauge sector
of QCD comes, for instance, from the temperature dependence of the
heavy quark potential. This is expected to change from a linearly rising
confinement potential to a Debye screened (Yukawa) potential at high
temperatures
$$
V(r,T) = \cases{
{-\alpha(T) \over r} + \sigma(T) r & $T < T_c$ \cr
{-\alpha (T) \over r} \exp{\{-m_D (T) r\}}& $T>T_c$ \cr}~~.\EQNO{potentt}
$$
The numerical studies
indicate that indeed the string tension decreases as one approaches $T_c$.
Furthermore, the confining part
of the potential disappears above $T_c$ and one is left with
a screened Coulomb potential. Some results for the screening mass,
$$
\mu  =\cases{
\sigma(T) /T & $T < T_c$ \cr
m_D (T) & $T > T_c$ \cr}~~.\EQNO{masst}
$$
are shown in \fig{potential} (Fukugita \etal~1989, Brown \etal~1988,
Bacilieri \etal~1988).
We note that at $T_c$ the correlation
length, $\xi = 1/\mu$,
in the pure $SU(3)$ gauge theory is $\simeq (1-2)T_c^{-1}$, \ie~
about (1-2)~fm. This shows that close to $T_c$ the system still is
strongly correlated.
However, as can be seen from \fig{potential},
already at temperatures $T \simeq 1.5 T_c$ the screening
mass is close to the leading order perturbative value
$$
m_D (T) =
\sqrt{{N\over 3} + {n_f\over 6}} g (T) T~~,\quad T>>T_c~~~.\EQNO{screen}
$$
In accordance with this perturbative result one finds that the
screening mass increases substantially, if dynamical quarks are present
in the medium (Karsch and Wyld 1988, Unger 1992).
This efficient screening of the heavy quark potential has important
consequences for the formation of heavy quark bound states in the
high temperature phase
(Karsch \etal~1988) which has been conjectured as a possible
signal for quark-gluon plasma formation in heavy ion collisions
(Matsui and Satz 1986).

\subsection{The Chiral Phase Transition}

Phase transitions are usually related to the breaking of global symmetries of
the Lagrangian. Besides a global $U(1)$ symmetry
corresponding to baryon number conservation the QCD
Lagrangian has, in the limit of vanishing quark masses,
a global chiral flavour symmetry, $SU_L(n_f) \times SU_R(n_f)$.
At low temperatures the chiral flavour symmetry is spontaneously broken
and it is expected, that it gets restored at high
temperatures. In the limit of vanishing quark masses this symmetry will control
the properties of the QCD phase transition. The expected critical
behaviour has been discussed some time ago by Pisarski and Wilczek (Pisarski
and Wilczek 1984).
On the basis of a renormalization group analysis of the most general 3-$d$
effective chiral Lagrangian they showed that the generic chiral transition is
expected to be first order as long as $n_f \ge 3$. In the most interesting case
of two massless flavours, however, the RG-analysis does not lead to a
decisive answer - a first as well as a second order transition is possible. In
the two flavour case the axial $U(1)$ symmetry, which at zero temperature is
broken due to the axial anomaly, plays an important role. If this symmetry
gets effectively restored at the chiral transition point, a first order
transition is possible also in the two flavour case. Otherwise the transition
would be second order with critical indices of the $O(4)$ Heisenberg model
(Pisarski and Wilczek 1984, Wilczek 1992).

Studies of the chiral transition on the lattice suffer from the fact
that one does not have a lattice formulation with the same symmetry
properties as the continuum action at hand. Most studies so far have
been performed with the staggered fermion action, which, for
$\bar{f}$ species of staggered fermions, is invariant only and a
$U (\bar{f}) \times U (\bar{f})$
subgroup of the the full flavour symmetry group.
Due to species doubling these $\bar{f}$
species of staggered fermions give rise to $n_f=4\bar{f}$ quark
flavours in the continuum limit. Only in this limit does one recover
a fermionic theory which has the correct $SU(4\bar{f}) \times SU(4\bar{f})$
chiral symmetry. The order parameter for this lattice chiral symmetry is
given by the chiral condensate,
$$
\langle \bar{\chi}\chi \rangle = {1 \over N_{\tau} N_{\sigma}^3}
{\partial \over \partial m} \ln Z~~.~~\EQNO{chiral}
$$

The importance of the different symmetry properties of the
lattice and continuum Lagrangian is evident in the case of $\bar{f}=1$,
which corresponds to
4-flavour QCD in the continuum limit. The lattice Lagrangian
in this case only has
a $U(1) \times U(1)$ symmetry, which is isomorphic to $O(2)$. One thus expects
a second order phase transition, with the critical exponents
of the 3-$d$ $O(2)$ spin
models as long as the flavour symmetry is strongly broken and the lattice
symmetry group  controls the dynamics. In the continuum limit,
on the other hand, one expects to find a first order phase transition.
This  change in critical behaviour has indeed been observed
in numerical simulations.
In the strong coupling limit the four flavour theory has a second order
phase transition
(Boyd \etal~1992a), while already at intermediate values of the
gauge coupling ($g^2 \simeq 1$) the transition is clearly first
order (Gupta \etal~1986, Karsch \etal~1987 and Gavai \etal~1990).

The four flavour theory has been studied for various values of the quark
mass. These studies suggest, that in the continuum limit there will be a
line of first order transitions connecting the zero-quark mass chiral
transition with the deconfinement transition in the limit $m \rightarrow
\infty$ (Gupta R \etal~1986). The situation for $n_f <4$ seems to be
different, although these studies can, at present, not be considered as
complete. In particular, one cannot rule out that the order of
transitions might change closer to the continuum limit, as we have
discussed above. The present understanding of the phase diagram for
three dynamical quarks with different masses is illustrated in the
generic phase diagram shown in \fig{generic} (Brown \etal~1990).
Whether the regions of first order transitions in the vicinity of the
chiral ($m_i=0$) and quenched ($m_i \rightarrow \infty$) limits are
really disjoint and whether the physically interesting case
of two light quarks, $m_u \simeq m_d \simeq 0$, and a heavier
strange quark, $m_s \simeq 150$~MeV, will lead to a first or second order
transition is currently subject of intensive research activities
(Petersson 1992).

Recently, first indications have been reported that a change in the
order of the transition might occur in two flavour QCD
(Mawhinney 1992 and Petersson 1992), similar to what has been found
in the four flavour case.
Due to this unsatisfactory situation it will be increasingly important
to reach consistent conclusions on the order of the transition by using
different lattice regularizations for the fermions.
Studies of the chiral transition with Wilson fermions are difficult, as the
lattice Lagrangian has no continuous chiral symmetry and the properties
related to the chiral flavour symmetry, which are so important
for the discussion
of phase transitions, can only be recovered in the continuum limit.
Consequently
the indications for a finite temperature phase transition with Wilson
fermions show, at least for small values of $N_{\tau}$
and thus at rather strong coupling, the features of
a continuous cross-over behaviour rather than a genuine phase transition
(Fukugita \etal 1986, Gupta R \etal 1989, Bitar \etal 1990b,1991).
Recent investigations of two flavour QCD with Wilson fermions
for $N_{\tau} =6$
suggest, however, a sharpening of the crossover behaviour,
when one comes deeper
into the regime of small lattice spacings. These studies even leave the
possibility for the existence of a first order phase transition
(Bernard etal 1992g).
They certainly have to be pursued in the future. The determination
of the transition temperature for Wilson as well as staggered
fermions also suggests
that with increasing $N_{\tau}$ the results extracted in both regularization
schemes start approaching each other.

A consequence of the spontaneously broken chiral symmetry in QCD is
the existence of a
massless Goldstone boson, the pion, as well as the non-degeneracy of
the
masses of chiral partners. These properties of the hadron spectrum
should change in the chirally symmetric high temperature phase. The
appropriate quantity to analyze on the lattice in this case is the large
distance behaviour of spacelike correlation functions for operators
with hadronic quantum numbers, $H(\s n,n_3)$ (DeTar and Kogut 1987a),
$$
  G_s^H(n_3,\s p) \; =\; {1\over N_{\tau} N_{\sigma}^3}
      \sum_{\s n=(n_0,n_1,n_2)} {\rm e}^{i\s p\cdot\s n}
        \langle H(\s n,n_3) H^\dagger(\s 0,0)\rangle\ ~~.\EQNO{Hscr}
$$
At large spatial separations these operators yield a screening
length, characteristic for the corresponding quantum number channels. As
long as there exist bound states in a given quantum number channel the
screening length will give the mass of the lowest lying state. We will
discuss this in more detail in the next subsection. In \fig{hscreen} we
show some results obtained in the staggered fermion formulation
for (pseudo)scalar and (pseudo)vector meson channels as
well as for the baryon channel
(DeTar and Kogut 1987a, b, Gottlieb \etal~1987c, Born
\etal~1991b, Boyd \etal~1992b).
The change in chiral properties of QCD is clearly visible in this
figure. All the chiral partners yield degenerate screening masses above
$T_c$. Similar results have been obtained with Wilson fermions
(Bitar \etal~1991).

\subsection{Spatial Correlations in the Quark-Gluon Plasma}

As discussed in section 4.1, the high temperature phase of QCD
is most certainly
not simply described by a weakly interacting gas of quarks and gluons. In
particular, in the vicinity of $T_c$ non-perturbative effects will modify the
spectrum strongly. We have discussed the effect on the equation of state and
the Debye screening lengths in the previous sections. A controversial issue is
the interpretation of so-called spatial correlation functions.
These are not readily related to physical
observables and (sometimes) show a behaviour which is not easily understood in
terms of ordinary high temperature perturbation theory.

The behaviour of the spatial hadronic correlation functions, defined
in \eq{Hscr}, can be analyzed in
terms of the spectral representation of these correlation functions.
At low temperatures the spectral functions have poles corresponding to
the (temperature dependent) masses of hadrons with the given quantum
numbers. In
the high temperature limit, however, the spectral function
may only have a cut,
corresponding to the free propagation of the minimal number of partons
in the hadronic channel considered, {\sl i.e.} two for mesonic and three
for baryonic operators. The cut will start at the lowest possible
momentum, $\s p_{\rm min}= (p_{0,{\rm min}}, 0, 0, 0,)$,
which due to the anti-periodic boundary conditions for the fermions
is given by the lowest Matsubara frequency,
$$
p_{0,{\rm min}} =\cases{
2\pi T &meson operators \cr
3\pi T &baryon operators \cr}~~.
\EQNO{matsubara}
$$
The screening masses for spatial hadronic operators thus are expected to
approach the lowest possible Matsubara frequency in the high temperature
limit, $\mu \sim p_{0,{\rm min}}$. This is indeed the case for the
baryonic, (pseudo)vector and also quark quantum number channels, as can
be seen in \fig{hscreen}. An exception is the
(pseudo)scalar screening mass. This clearly deviates
strongly from
the free field behaviour and also cannot be understood in terms of small
perturbative corrections. Investigations of the quark mass dependence of
the scalar screening mass suggest, however, that these screening mass is
not associated with a bound state but rather reflects the still rather
strong attractive interactions in this channel (Gupta S
1992b).

Spatial hadronic correlation functions are in so far rather simple as
their high temperature behaviour can be studied and understood in terms
of high temperature perturbation theory. This is not the case for the
behaviour of other spatial observables like spatial Wilson loops or
spatial four-point correlation functions. The former still show area
law behaviour in the high temperature phase (Borgs 1985 and Manousakis
and Polonyi 1987) and thus give rise to a non-perturbative, spatial
``string tension". The occurrence of such a string tension can be
understood by means of dimensional reduction (Appelquist and Pisarski
1981) of a quantum field theory at high temperature. For QCD the
integration over static modes leads, in the high temperature limit,
to an adjoint Higgs model in three dimensions which shows confining
properties for the spatial gauge degrees of freedom, although the potential
between the timelike gauge fields (Higgs fields) is screened. The short
distance properties of this effective model have been studied in
perturbation theory (Reisz 1991, 1992 and Lacock \etal 1992) and can
be related to the short distance properties of QCD at high temperatures.
However, in how far the large distance properties of the effective
theory describe the large distance properties of QCD, is at present
unknown and deserves more detailed studies.
As the non-static
modes do not decouple completely in the high temperature limit (Landsman
1988), it is to be expected that
the effective theory becomes increasingly complicated,
if it should
describe also the large distance properties of QCD.

Even less understood is, at present, the behaviour of point-splitted
quark four-point functions,
which at zero temperature are used to study hadronic wave-functions. At
finite temperature their physical interpretation is not at all obvious
(Bernard \etal~1992f). The spatial ``wave-functions"  calculated for
various quantum number channels still are strongly localized while leading
order perturbation theory would have suggested very broad ``wave-functions".
It is not clear whether these calculations indicate that there are
localized ``quasi"-particles with
the quantum numbers of e.g. a pion or a baryon in the high
temperature phase.
Most likely this is not the case. Various measurements
of physical quantities like
the baryon number susceptibility (Gottlieb \etal~1987d and 1988)
or the baryon number distribution
in the vicinity
of static quark sources
(Bernard \etal~1992h)
suggest that quarks can move quite freely in the
high temperature phase.

\section{Electro-weak sector of the Standard Model}

The Glashow-Salam-Weinberg theory is very successful in describing
electro-weak interactions. Experimentally, there are no signs of its
breakdown. Still, from a theoretical point of view,
the theory is not satisfactory. Besides the number of free
parameters being large, neither the Higgs sector nor
the $U(1)$ gauge subgroup present asymptotically free theories.
This indicates their breakdown at large energies (couplings).
One of the main motivations
for lattice studies of the standard model has therefore been
to possibly quantify its limitations as a consistent quantum field theory
by analyzing it at a non-perturbative level or to detect deviations
from the perturbative Callan-Symanzik $\beta$-functions which would
lead to a different behaviour in the large cut-off limit.

\subsection{Pure Higgs Systems}

In the electro-weak sector of the standard model,
the $SU(2)_L \times U(1)_Y$ gauge couplings are small at the
$m_W$ scale. Likewise, except for a very large top quark mass,
the Yukawa couplings of the quarks are also small. Thus, at least
in a first approximation, one may ignore gauge fields and fermions
and concentrate on the Higgs sector alone. This is then an
$O(4)$ invariant scalar field theory,
$$
{\cal L} = {1\over 2}(\partial_{\mu} \Phi)^2 + {1\over 2}m^2_0 \Phi^2
+ {g_0 \over{4!}}\Phi^4 .
\EQNO{eqhiggs}
$$
In a (semi)classical treatment, the Higgs-mechanism
is based on the observation that for $m^2_0 \leq 0$
a minimum value for the action is obtained when the Higgs field acquires
a non-vanishing value. This leads, in the unitary gauge, to a
gauge boson mass of
$$
m_W^2 = \frac{1}{4} g_L^2 \langle \Phi \rangle^2 ,
\EQNO{aaa}
$$
where $g_L$ is the renormalized $SU(2)_L$ gauge subgroup coupling
and $\langle \Phi \rangle$
represents the Higgs field expectation value in the vacuum.
This relation remains valid for arbitrarily large Higgs
self-couplings $g$ as long as the gauge coupling can be
treated perturbatively (Dashen and Neuberger 1983).
On the other hand, the mass of the Higgs
boson is given by
$$
m_H^2 = {1\over 3} g_R \langle \Phi \rangle^2 .
\EQNO{eqhiggsmass}
$$
Here, the renormalized quartic coupling $g_R$ enters crucially.
In particular, by means of a combination of strong-coupling expansion
and renormalization techniques it was shown (L\"uscher and Weisz 1987,
1988, 1989)
that in $4$ dimensions scalar field theories are trivial,
i.e. the renormalized
quartic coupling vanishes when the ultraviolet cut-off is taken
to infinity. Further investigations at small gauge coupling,
perturbative calculations
as well as numerical lattice simulations, confirmed that the triviality
is not destroyed when $SU(2)$ gauge fields are reintroduced
(Hasenfratz A and Hasenfratz P 1986, Langguth and Montvay 1987,
Hasenfratz A and Neuhaus 1988, Bock \etal 1990a).
A non-trivial fixed point at intermediate
values for the gauge coupling is not entirely excluded
but its existence seems unlikely because the Higgs phase-transition
line appears to be first order everywhere, in agreement with
the original work of Coleman and Weinberg
(Coleman and Weinberg 1973).
Thus, the standard model should be the effective
low-energy limit of some larger theory, valid
below an energy scale set by an intrinsic cut-off $\Lambda$.
Equation \eq{eqhiggsmass} then implies an upper bound on the Higgs mass:
a given value for the momentum cut-off leads to a maximum value
for the coupling $g$ which decreases further when $\Lambda$ is
increased. The quantitative determination of the bound on
the Higgs mass can in principle
require a non-perturbative treatment because $g$ needs not to be
small for small values of the cut-off. This question has been addressed
both by analytical calculations (L\"uscher and Weisz 1987, 1988,
1989)
and numerical simulations (Hasenfratz A \etal 1987, Kuti \etal 1988).
The analysis
has been refined since then, in particular in regard to
finite-size effects due to the presence of massless Goldstone-bosons
in the symmetry breaking phase
(Neuberger 1988, Hasenfratz P and Leutwyler 1990, Hasenfratz A \etal 1990,
G\"ockeler \etal 1992a).
Some results are shown in \fig{fighiggsmass}.
Moreover, the universality of
the upper bound has been investigated \ie~the dependence of the
Higgs mass on the specific lattice discretization scheme has been checked
(Bhanot \etal 1990, G\"ockeler \etal 1992a).
Differences of $m_H / \langle \phi \rangle$
between various lattice regularization schemes are not too surprising as the
cut-offs in physical units are different. Moreover, the finite
result for the upper bound depends on the
physical criterion which specifies how much
cut-off dependence is allowed in what physical quantity.
A popular choice has been the somewhat ad hoc condition
$\Lambda > 2 m_H$. Another procedure utilizes the
$90^{\circ} \; W W$ scattering cross section
(Bhanot \etal 1990, G\"ockeler \etal 1992a).
Correspondingly,
the results exhibit
a $10 - 20 \%$ variation. Nevertheless it is quite clear
that a Higgs mass larger than $600(60)$~GeV cannot be accommodated
by the standard model. This number is consistent with the
perturbative unitarity bound of $m_H < 800$~GeV
(Lee \etal 1977).
The reason is
that even for a low cut-off, $\Lambda = m_H$, the renormalized
quartic Higgs self-coupling is not large. Yet it is reassuring,
that the perturbative estimate is under non-perturbative control.

\subsection{The Higgs Phase Transition}

The spontaneously broken symmetry in the Higgs field sector of the
standard model is expected to be restored at high temperature
(Kirzhnits and Linde 1972, Weinberg 1974).
The order parameter for this phase transition is the
expectation value of the Higgs field, $\langle \Phi \rangle$, which at
zero temperature takes on the value, $\langle \Phi \rangle \simeq 250$~GeV.
The perturbative calculation of the effective
Higgs potential becomes rather involved, once higher order corrections as well
as the influence of the Higgs and W-Boson masses are taken into account
(Dine \etal 1992, for a recent review see also Kripfganz 1992). The
transition turns out to be first order with a critical temperature given
by,
$$
T_{\rm SR} \simeq \sqrt{{2\over 3}} \langle \Phi \rangle{m_H \over m_W}~~.
\EQNO{tchiggs}
$$
The transition is weakly first order as the discontinuity in the order
parameter turns out to be much smaller than the transition
temperature itself,
$$
\langle \Phi \rangle (T_{\rm SR}) \simeq  {m_W^3 \over m_H^2
\langle \Phi \rangle (0)} T_{\rm SR}~~.
\EQNO{pchiggs}
$$

The occurrence of a symmetry restoring phase transition has been
established in numerical simulations of the $SU(2)$-Higgs model at
finite temperature (Evertz \etal 1987, Damgaard and Heller 1988) and
the recent calculations in the region of small quartic Higgs coupling
(Bunk \etal 1992) suggest that this transition indeed is first order.
These latter calculations aim, for the first time, at a quantitative
determination of the critical temperature in a parameter regime where
the Higgs boson mass is approximately equal to the W-boson mass. It has been
found that the discontinuity in the Higgs condensate at $T_{\rm SR}$ is
about twice as large as expected from perturbation theory,
$\langle \phi \rangle_R (T_{\rm SR})\simeq 0.68 T_{\rm SR}$. The critical
temperature itself has been estimated as
$$
T_{\rm SR} \simeq 1.74~m_W \simeq 140~{\rm GeV}.
\EQNO{SRtemp}
$$

In the absence of gauge degrees of freedom, \ie~in the $O(4)$ symmetric
Higgs sector defined by \eq{eqhiggs}, the Higgs phase transition has been
analyzed numerically using finite size scaling techniques
(Jansen and Seuferling 1990).
Here it has been found that the critical parameter of the transition are
in good quantitative agreement with predictions based on
renormalized perturbation theory (L\"uscher and Weisz 1989).
This should not be too surprising as the studies at zero temperature
have indicated that
the renormalized quartic Higgs coupling never becomes large. Consequently the
perturbative estimates for the Higgs phase transition are also
expected to be rather accurate. In the $O(4)$ model one
finds for the critical temperature $T_{\rm SR} \simeq \sqrt{2} \langle
\phi \rangle_R$. This is consistent with a determination of a lower
bound for the symmetry restoration temperature (Gavai \etal 1992),
which can be deduced
from the upper bound for the Higgs boson mass, discussed in the previous
section.
One finds as a lower bound
$$
{T_{\rm SR,min} \over m_H} = 0.58 \pm 0.02~~,
\EQNO{tcbound}
$$
when the Higgs mass reaches the cut-off, \ie~$m_Ha =0.5$.
Using for the Higgs mass
the upper bound $m_H < 600$~MeV, one thus obtains
$T_{\rm SR,min}=350$~GeV as a lower
bound for the symmetry restoration temperature in the $O(4)$ model.
This is more than a factor two larger than the value found for the transition
temperature in the simulation of the $SU(2)$ Higgs model.
In the future it thus
will be important to understand in more detail the
dependence of $T_{\rm SR}$ on
the parameters of the model, in particular its dependence
on $m_H$ and $m_W/m_H$.

\subsection{Higgs-Yukawa Couplings}

Higgs-Yukawa couplings of heavy quarks to the Higgs field
are interesting for several reasons.
The presence of heavy quarks could affect the upper bound on
the Higgs mass. Secondly,
from the triviality of Higgs-Yukawa couplings,
upper bounds on heavy quark masses can be derived
much like in pure Higgs systems.
Finally, from a general field theoretical point of view,
it is a challenge to put chiral fermions on the lattice.
Phenomenologically, this is of principal importance
if there is a strongly interacting sector with chiral
fermions in theories with larger symmetry groups than in the standard model.

The problem of treating chiral fermions in the lattice
regularization
consists in the occurrence of the so-called doubler fermions.
If one discretizes the Dirac action additional fermions
are generated such that the number of left- and right-handed
fermions is equal. As mentioned in section 2.1 this is a general
phenomenon. There are, however,
certain loopholes. For example,
in the Wilson discretization of fermions, \eq{eqwilsonferm},
a second-order derivative term $S_W$
which becomes irrelevant in the (naive) continuum limit
was added to the naively discretized fermion action $S_F$.
For the doubler fermions, this term generates a
contribution to the mass which is
proportional to $1/a$ so that they acquire infinite
mass in the continuum limit $a \rightarrow 0$ and decouple.
As this extra term couples left-handed to right-handed
fermions chiral invariance is lost.

A number of proposals to overcome this difficulty
have been suggested in the last few years.
Quite generally, the situation is rather complex.
For each model which typically depends on quite a few
parameters one
has to map out the phase diagram.
A careful analysis of the
usually rich phase structure is then necessary
to locate the region(s) in parameter space
where a sensible continuum limit can be taken.
All this requires a lot of numerical and analytical work
and we cannot even attempt to present a comprehensive
overview over all the results obtained so far.
The material is discussed in much greater detail
in recent reviews
(De and Jersak 1992, Montvay 1992) to which
we refer also for a more complete and fair list of references.
Here we can only sketch some basic ideas and difficulties
in this field,

A rather intensively investigated approach to chiral fermions
on the lattice
is the Smit-Swift model (Smit 1980, 1986, Swift 1984).
The action of this model consists of
a standard Higgs part $S_H$ (the lattice equivalent to
\eq{eqhiggs}),
the naive fermion action $S_F$, \eq{eqwilsonferm},
a Yukawa term $S_Y$ plus
a so-called Wilson-Yukawa part $S_{WY}$:
$$
S = S_H + S_F + S_Y + S_{WY}
\EQNO{Rafaela}
$$
with
$$
\eqalign{
S_Y & =   y \sum_n \{ \psib_L(n) \phi(n) \psi_R(x)
                    + h.c. \}
          \cr
S_{WY} & =   {w\over 2} \sum_n \{
       2  \psib_L(n) \phi(n) \psi_R(n) \cr
  &   -  \psib_L(n) \phi(n) \psi_R(n+\hat{\mu})
       -  \psib_L(n+\mu) \phi(n+\hat{\mu}) \psi_R(n)
         + h.c. \}  \cr}
\EQNO{Carino}
$$
Here, $\psi_{L,R}$ denote the chiral projections ${1\over 2}
(1 \mp \gamma_5) \psi$. The action is invariant under the global
chiral $SU(2)_L \times SU(2)_R$ transformations
$$
\eqalign{
\psi_L \rightarrow \Omega_L \psi_L &,  \psib_L \rightarrow \psib_L
\Omega_L^{\dagger} \cr
\psi_R \rightarrow \Omega_R \psi_R &,  \psib_R \rightarrow \psib_R
\Omega_R^{\dagger} \cr
\phi \rightarrow \Omega_L \phi \Omega_R^{\dagger} & \cr}
\EQNO{Dux}
$$
where
$\Omega_L \in SU(2)_L$ and $\Omega_R \in SU(2)_R$.
\footnote{$^{\dagger}$}
{The model has also been analyzed with simpler symmetry groups
as e.g. $U(1)_L \times U(1)_R$.}
The $SU(2)_L$ symmetry group is to be gauged later in order to make
contact with the standard model. First of all, one has to deal with the
doubler fermions half of which come with the opposite ``mirror"
behaviour under chiral transformations. By analyzing the
fermion propagator it could be shown that at strong
Wilson-Yukawa coupling $w$ the doublers acquire a mass which stays
at order $1/a$ when the (bare) couplings are tuned to
certain critical values, \ie~close
to one of the second order phase transitions
(Thornton 1989, Bock \etal 1989, 1990b, Smit 1989,
Golterman and Petcher 1990, Aoki \etal~1990).
Thus, there is a region in the space of parameters where
the doublers are removed in the continuum limit.
Still, the theory contains a right-handed neutrino as an elementary
field. It could, however, be proven that
at $y = 0$ the action possesses
a shift symmetry in the right-handed fields
(Golterman and Petcher 1989).
Due to this symmetry the right-handed neutrino decouples,
even when gauge interactions are switched on again.
Both results had made the Smit-Swift model a promising
candidate to investigate the standard model on the lattice.
There is, however, a serious drawback.
Apart from the elementary fields one has to analyze
the spectrum of the model.
In particular, there are composite
mirror fields like $\phi^{\dagger} \psi_L$
which are left-handed but transform non-trivially under
$SU(2)_R$. The propagation of such
composite fermion operators has been looked at and
the corresponding masses have been determined
(Bock and De 1990, Bock \etal 1992).
It turns out that a Dirac fermion,
$\phi^{\dagger} \psi_L + \psi_R$,
which is neutral under $SU(2)_L$ but transforms vectorially under
$SU(2)_R$ remains in the theory.
This is of course not acceptable in the standard
model.

As the presence of mirror fields apparently cannot be avoided,
it has been suggested to include them as fundamental
fields (Montvay 1987a, b). This proposal offers the opportunity
of more flexibility and also of better control
over the decoupling of mirror fermions.
The fermion fields $\psi$ are
supplemented by mirror fields $\chi$ such that $\chi_R$
transforms the same way as $\psi_L$. Then one can write
chirally invariant mixed terms like $\psib_L \chi_R$ etc..
The Wilson term is designed to offer the possibility of
removing both fermion and mirror doublers,
$$
\tilde{S}_W =
        {r\over 2} \sum_{n,\mu}
                \{ 2 \psib(n) \chi(n)
                   - \psib(n+\hat{\mu}) \chi(n)
                   - \psib(n) \chi(n+\hat{\mu})
                   + ( \psi \leftrightarrow \chi ) \},
\EQNO{Fredo}
$$
while there are separate Yukawa terms for $\psi$ and $\chi$
$$
\eqalign{
\tilde{S}_Y  &=   y_{\psi} \sum_n \{ \psib_L(n) \phi(n) \psi_R(n)
                    + h.c. \} \cr
             &+   y_{\chi} \sum_n \{ \chib_L(n) \phi^{\dagger}(n)
                                      \chi_R(n) + h.c. \} .\cr}
\EQNO{Westbury}
$$
In addition, a mixed bare mass term
$\sum_n m \{ \psib(n) \chi(n) + h.c. \}$
is present but no Wilson-Yukawa part.
The model has mainly been investigated for
a global chiral $U(1)_L \times U(1)_R$ symmetry. There are
more free parameters than in the Smit-Swift model so that
mapping out the phase diagram is even more demanding.
Still, numerical results indicate that the doublers can be
decoupled (Lin \etal~1990, 1991a, Farakos \etal~1991).
However, the Dirac fields
$\psi_A = \psi_L + \chi_R$ and $\psi_B = \psi_R + \chi_L$
transform non-trivially only under $U(1)_L$ and $U(1)_R$
respectively in the symmetric phase, which is completely vector-like
when written in those fields,
while in the broken phase they transform equally.
Moreover, the shift symmetry is of no help when
gauge interactions are turned on so that presumably
the mirror fermions do not decouple.
The real question is therefore to what extent the degeneracy
between fermions and mirror fermions can be lifted in the broken phase
in order
not to come into conflict with phenomenology. This
again requires careful investigations
in various regions of parameter space,
a formidable numerical task
when it is attempted to derive quantitative results.
There are indications that by
a fine tuning of parameters the ratio of mirror fermion mass
to the Higgs condensate, $m / \langle \phi \rangle$,
can be made
as large as 6, corresponding to a mirror fermion mass in
the TeV range (Lin \etal~1991b).
Still, the properties of the model in the
symmetric phase are probably not satisfactory for an
understanding of symmetry breaking in theories with chiral fermions.

To end this section we mention two other approaches. In
(Borelli \etal 1990) the formulation of a chiral gauge theory
without mirror fermions is based on gauge fixing. The second proposal
attempts to obtain chiral fermions as zero modes bound to a domain wall.
This is generated, in five dimensions, by a mass term which depends
on the extra dimension and which has the form of a kink (Kaplan 1992).
The analysis of these ideas is not as advanced as in the
Smit-Swift or the mirror model and it remains to be seen whether
one can achieve a thorough understanding of the chiral sector
of the standard model.

\subsection{Strong-coupling QED}

QED is the best tested of all field theories and it describes,
at currently explorable energy scales,
the interactions of charged particles with
remarkable precision.
Yet, from perturbation theory
it is known that for all finite
values of the bare coupling the renormalized coupling goes to zero
if the ultraviolet cut-off is sent to infinity - the so-called
Landau pole.
Even if this remains true for the exact solution,
pure QED would still be a useful low-energy effective theory.
However, at large
energies the effective charge becomes large
so that perturbation theory should not be trusted anymore.
It is therefore possible
that through non-perturbative effects the Callan-Symanzik
$\beta$-function acquires an ultraviolet stable fixed point
at which the theory could develop non-trivial new physics.
For pure QED this problem might be regarded as academic,
as the
cut-off can be pushed to values beyond the Planck scale where
QED should not be considered in isolation anyway.
Nevertheless, in the context of more general models
non-perturbative phenomena of non-asymptotically free theories
may play an important role and QED provides a useful laboratory
to study those effects.

A renewed interest in non-perturbative studies of QED arose after
an investigation of the truncated Schwinger-Dyson equation
for the fermion propagator revealed a continuous chiral phase
transition, with chiral symmetry being broken spontaneously
at strong coupling (Miransky 1985).
Although the calculation did not include
vacuum polarization effects it was argued that the critical coupling
should be regarded as an ultraviolet stable fixed point at which
the theory admits a non-trivial continuum limit. The existence
of a chiral transition was confirmed by numerical studies
of non-compact lattice QED (Kogut \etal 1988, 1989).
These lattice investigations also claimed to find support
for a non-trivial critical behaviour. Later on, however, it was found
(Booth \etal 1989, Horowitz 1990, G\"ockeler \etal 1990a)
that, at least for
four flavours, the critical exponents are consistent with
mean field behaviour.
In particular,
the numerical computation of the Callan-Symanzik $\beta$ function
for the renormalized charge (G\"ockeler \etal 1990b) showed that the charge
matches to one-loop renormalized perturbation theory
and vanishes in the continuum limit.
In the matter sector,
the scaling behaviour was also found to be
consistent with triviality
(G\"ockeler \etal 1992a).
This corresponds to a large anomalous
dimension for the chiral condensate,
$\gamma_{{\overline \chi} \chi} = - 2$,
which also indicates renormalizability
of a four-fermion interaction (Booth \etal 1989).
Analytically, these observations are
supported by studies of a coupled set of Schwinger-Dyson equations
which include certain effects of fermion
loops
(Rakow 1991, Kondo 1991).
These and accompanying results suggest that
the chiral transition of QED and of
the Nambu-Jona Lasinio model
are in the same universality class (Horowitz 1990).

It must be said, however, that these findings are not yet completely
settled. The main issues being debated are the exact form
of the scaling laws (Kocic \etal 1992a),
a problem which will require a large
computational effort to be unambiguously clarified, and the
occurrence (Kocic \etal 1992b, Hands \etal 1992, Kocic \etal 1992c)
or not (Rakow 1992) of a monopole phase transition. If the chiral
transition is driven by monopole condensation it would be a
lattice artifact. Otherwise,
in view of the large anomalous dimensions being observed,
it may be concluded that non-asymptotically free theories can posses
interesting new features not covered by perturbation theory.

\section{Outlook}

Numerical simulations on space-time lattices to date offer the most
promising prospects to analyze non-perturbative phenomena
in the standard model of electro-weak and strong interactions
and beyond.
In this review we have presented in some detail the
numerical techniques which have
been developed over the last ten years and which are
now being applied in large scale numerical calculations.
For
basic questions like the QCD mass spectrum or the heavy quark potential
these calculations provide answers, which can and have
successfully been compared
with existing experimental data. Other calculations like those for
the QCD phase transition or the weak matrix elements will in the near
future reach an accuracy where they can make predictions for new
experiments.

Nevertheless, there are open questions within the
standard model which could not yet be addressed satisfactorily by
present day methods. For instance, in questions like the role of
$\Theta$-vacua or QCD at non-vanishing baryon number density
one is led to
path integrals with complex actions. So far no satisfactory algorithm
has been found for such problems (for a review see e.g. Barbour 1992).
The analysis of the electro-weak sector is still incomplete.
One has not yet succeeded in developing
a convincing formulation of chiral
fermions on the lattice. New suggestions still have to be explored.
In this sector, there is also considerable interest in
understanding the temperature dependence of
baryon number violating processes, which play a central role in
the generation of a baryon number asymmetry during the
electro-weak phase transition (for a recent review see e.g. Shaposhnikov 1992).

The study of the gravitational force or, more generally, string models on
the lattice has not been mentioned by us at all. Here one
usually follows quite a different approach. Rather than formulating
the theory on a rigid hypercubic lattice one constructs dynamically
triangulated lattices, which define random surfaces (for a recent review
see e.g. Kawai 1992).
Present investigations concentrate on understanding the phase structure
of random surface models and on the identification of critical points at
which a continuum limit can be taken. Not surprising, also here the
inclusion of fermions is of central interest.

In the future, we expect that
the numerical analysis of quantum field
theories will even more so be a valuable tool to study quantitatively
non-perturbative aspects of these theories, which otherwise would
not be accessible. Although the analysis of many questions of interest
is still limited by the available computer resources progress in the
field is very rapid. It is to be expected that within 2-3 years
the basic questions in QCD will be investigated on
massively parallel computers,
which will deliver more than $10^{12}$ floating point operations per second
(Teraflops). This will exceed the presently available resources for
typical QCD-projects by two orders of magnitude. Parallel to these
hardware improvements the algorithms used in the simulation of quantum
field theories have much improved. One thus can expect that in the
future numerical techniques will be applied to questions which
up to now are considered to be too difficult (Parisi
1992) computationally.

\ack
We wish to thank Sourendu~Gupta, U.~Heller,
R.~Horsley, M.~Laursen and Th.~Neuhaus for
discussions and for providing material
for the preparation of this article.
This
work was partially supported by the DFG under grant PE 340/3-1.

\references

\refjl{Abada A \etal 1992} {\NP}{B376}{172}
\refjl{Adler S L 1981}{\PR}{D23}{2901}
\refjl{Adler J \etal (MARK III Collaboration) 1989}{\PRL}{62}{1821}
\refjl{Albanese M \etal (APE Collaboration) 1987}{\PL}{B192}{163}
\refjl{Alexandrou C 1991} {\PL}{B256}{60}
\refjl{Allton C R \etal 1991}{\NP}{B349}{598}
\refjl{Allton C R \etal (UKQCD Collaboration) 1992a}{\PL}{B284}{377}
\refjl{Allton C R \etal (UKQCD Collaboration) 1992b}{\PL}{B292}{408}
\refjl{Altmeyer R \etal ($MT_c$ Collaboration) 1993}{\NP}{B389}{445}
\refjl{Alvarez O 1981}{\PR}{D24}{440}
\refjl{Anjos J \etal (E691 Collaboration) 1989}{\PRL}{62}{1587}
\refjl{Anjos J \etal (E691 Collaboration) 1990}{\PRL}{65}{2630}
\refjl{Aoki S \etal 1990}{\PL}{B243}{403}
\refjl{Appelquist T and Pisarski R D 1981}{\PR}{D23}{2305}

\refjl{Bacilieri P \etal (APE Collaboration) 1988}{\PRL}{61}{1545}
\refjl{Bali G and Schilling K 1992}{\PR}{D46}{2636}
\refjl{Bali G and Schilling K 1993}{\PR}{D47}{661}
\refjl{Barbour I 1992}{\NP (Proc. Suppl.)}{B26}{22}
\refjl{Batrouni G G \etal 1985}{\PR}{D32}{2736}
\refjl{Bernard C 1974}{\PR}{D9}{3312}
\refjl{Bernard C \etal 1988}{\PR}{D38}{3540}
\refjl{Bernard C and Soni A 1990}{\NP (Proc. Suppl.)}{B17}{495}
\refjl{Bernard C \etal 1991}{\PR}{D43}{2140}
\refjl{Bernard C \etal 1992a}{\PR}{D45}{869}
\refjl{Bernard C \etal 1992b}{\NP (Proc. Suppl.)}{B26}{204}
\refjl{Bernard C \etal (MILC Collaboration) 1992c}
      {\NP (Proc. Suppl.)}{B26}{262}
\refjl{Bernard C \etal (MILC Collaboration) 1992d}
      {\NP (Proc. Suppl.)}{B26}{305}
\refjl{Bernard C \etal 1992e}{preprint UW/PT-92-21}{}{}
\refjl{Bernard C \etal (MILC Collaboration) 1992f}{\PRL}{68}{2125}
\refjl{Bernard C \etal (MILC Collaboration) 1992g}{preprint AZPH-TH/92-16}{}{}
\refjl{Bhanot G \etal 1990}{\NP}{B353}{551}
\refjl{Billoire A 1980}{\PL}{B92}{343}
\refjl{Billoire A \etal 1985}{\NP}{B251}{581}
\refbk{Binder K 1979}{Monte Carlo Methods in Statistical Physics}{
      Springer, Berlin}
\refjl{Bitar K M \etal (HEMCGC Collaboration) 1990a}{\PR}{D42}{3794}
\refjl{Bitar K M \etal (HEMCGC Collaboration) 1990b}{\PL}{B234}{333}
\refjl{Bitar K M \etal (HEMCGC Collaboration) 1991}{\PR}{D43}{2396}
\refjl{Bitar K M \etal (HEMCGC Collaboration) 1992}
      {preprint FSU-SCRI-92-152}{}{}
\refjl{Bock W \etal 1989}{\PL}{B232}{486}
\refjl{Bock W and De A K 1990}{\PL}{B245}{207}
\refjl{Bock W \etal 1990a}{\PR}{D41}{2573}
\refjl{Bock W \etal 1990b}{\NP}{B344}{207}
\refjl{Bock W \etal 1992}{\NP}{B388}{243}
\refjl{Booth S \etal 1989}{\PL}{B228}{115}
\refjl{Booth S \etal (UKQCD Collaboration) 1992}{\PL}{B294}{385}
\refjl{Borelli A \etal 1990}{\NP}{B333}{335}
\refjl{Borgs C 1985}{\NP}{1985}{455}
\refjl{Born K D \etal ($MT_c$ Collaboration) 1991a}
        {\NP (Proc. Suppl.)}{B20}{394}
\refjl{Born K D \etal 1991b}{\PRL}{67}{302}
\refjl{Boyd G \etal 1992a}{\NP}{B376}{199}
\refjl{Boyd G \etal 1992b}{\NP}{B385}{481}
\refjl{Bowler K \etal 1986}{\PL}{B179}{375}
\refjl{Brown F R and Woch T J 1987}{\PRL}{58}{2394}
\refjl{Brown F R \etal 1988}{\PRL}{61}{2058}
\refjl{Brown F R \etal 1990}{\PRL}{65}{2491}
\refjl{Brown F R \etal 1991} {\PRL}{67}{1062}
\refbk{Buchm\"uller W and Cooper S 1988 in}{High Energy Physics}{
      Ali A and S\"oding P (eds), World Scientific, Singapore}
\refjl{Bunk B \etal 1992}{\PL}{B284}{371}
\refjl{Butler F \etal (GF11 Collaboration) 1992}{in preparation}{}{}

\refjl{Cabasino S \etal (APE Collaboration) 1991}{\PL}{B258}{202}
\refjl{Cabibbo N and Marinari E 1982}{\PL}{B119}{387}
\refjl{Callaway D J E and Rahman A 1982}{\PRL}{49}{613}
\refjl{Campostrini M \etal 1986}{\PRL}{57}{44}
\refjl{Campostrini M and Rossi P 1990}{\NP}{B329}{753}
\refjl{Caracciolo S \etal 1992}{preprint FSU-SCRI-92-65}{}{}
\refbk{Christ N H 1990 in}{Proceedings of the 8th Conference on
       Computing in High Energy Physics}{Santa Fe}
\refjl{Coleman S and Weinberg E 1973}{\PR}{D7}{1888}
\refjl{Creutz M \etal 1979}{\PRL}{42}{1390}
\refjl{Creutz M 1980}{\PR}{D21}{2308}
\refjl{Creutz M 1987}{\PR}{D36}{515}
\refjl{Creutz M 1988}{\PR}{D38}{1228}
\refbk{Creutz M (ed) 1992}{Quantum Field Theory on the Computer}{
       Advanced Series on Directions in High Energy Physics,
       World Scientific, Singapore}
\refjl{Damgaard P H and Heller U M 1988}{\NP}{B304}{63}
\refjl{Daniel D \etal 1992} {\PR}{D46}{3130}
\refjl{Dashen R and Neuberger H 1983}{\PRL}{50}{1897}
\refbk{De A K and Jersak J 1992 in}{Heavy Flavours}{Buras A J and
       Lindner M (eds), World Scientific, Singapore}
\refjl{DeTar C and Kogut J 1987a}{\PRL}{59}{399}
\refjl{DeTar C and Kogut J 1987b}{\PR}{D36}{2828}
\refjl{DeTar C 1988}{\PR}{D37}{2328}
\refjl{Dine M \etal 1992}{\PR}{D46}{550}
\refjl{Drouffe J M and Itzykson C 1978}{Phys. Rep.}{38}{133}
\refjl{Duane S \etal 1987}{\PL}{B195}{216}

\refjl{Eichten E \etal 1975}{\PRL}{34}{369}
\refjl{Eichten E and Feinberg F 1981}{\PR}{D23}{2724}
\refjl{Eichten E 1988} {\NP (Proc. Suppl.)}{B4}{70}
\refjl{El-Khadra A X 1992}{\NP (Proc. Suppl.)}{B26}{372}
\refjl{El-Khadra A X \etal 1992}{\PRL}{69}{729}
\refjl{Engels J \etal 1990a}{\PL}{B252}{625}
\refjl{Engels J \etal 1990b}{\NP}{B332}{737}
\refjl{Evertz H G \etal 1987}{\NP}{B285}{229}

\refjl{Farakos K \etal 1991}{\NP}{B350}{474}
\refjl{Fingberg J \etal 1992}{preprint BI-TP 92/26}{}{}
\refjl{Fischler W 1977}{\NP}{B129}{157}
\refjl{Fucito F \etal 1981}{\NP}{B180}{369}
\refjl{Fukugita M and Ukawa A 1985}{\PRL}{55}{1854}
\refjl{Fukugita M \etal 1986}{\PRL}{57}{1974}
\refjl{Fukugita M \etal 1989}{\PRL}{63}{1768}
\refjl{Fukugita M \etal 1990}{\NP}{B337}{181}
\refjl{Fukugita M \etal 1992a}{\NP (Proc. Suppl.)}{B26}{284}
\refjl{Fukugita M \etal 1992b}{\NP (Proc. Suppl.)}{B26}{265}
\refjl{Fukugita M \etal 1992c}{preprint KEK TH-340}{}{}

\refjl{Gasser J and Leutwyler H 1982}{Phys. Rep.}{87}{77}
\refjl{Gavai R V \etal 1990}{\PL}{B241}{567}
\refjl{Gavai R V \etal 1992}{\PL}{B294}{84}
\refjl{Gavela M B \etal 1988a}{\PL}{B206}{113}
\refjl{Gavela M B \etal 1988b}{\NP}{B306}{677}
\refjl{G\"ockeler M \etal 1990a}{\NP}{B334}{527}
\refjl{G\"ockeler M \etal 1990b}{\PL}{B251}{567}
\refjl{G\"ockeler M \etal 1992a}{preprint HLRZ 92-35}{}{}
\refjl{G\"ockeler M \etal 1992b}{\NP}{B371}{713}
\refjl{Goltermann M F L and Petcher D N 1989}{\PL}{B225}{159}
\refjl{Goltermann M F L and Petcher D N 1990}{\PL}{B247}{370}
\refjl{Gottlieb S \etal 1987a}{\PR}{D35}{2531}
\refjl{Gottlieb S \etal 1987b}{\PRL}{59}{1881}
\refjl{Gottlieb S \etal 1988a}{\PR}{D38}{2245}
\refjl{Gottlieb S \etal 1988b}{\PR}{D38}{2888}
\refjl{Gottlieb S \etal 1992}{\NP (Proc. Suppl.)}{B26}{308}
\refjl{Gromes D 1984}{\ZP}{C26}{401}
\refjl{Guagnelli M \etal (APE Collaboration) 1990}{\PL}{B240}{188}
\refjl{Guagnelli M \etal (APE Collaboration) 1992}{\NP}{B378}{616}
\refjl{Gupta R \etal 1986}{\PRL}{57}{2621}
\refjl{Gupta R \etal 1988a}{\PL}{B211}{132}
\refjl{Gupta R \etal 1988b}{\PR}{D38}{1278}
\refjl{Gupta R \etal 1989}{\PR}{D40}{2072}
\refjl{Gupta R \etal 1991a} {\PR}{D43}{2003}
\refjl{Gupta R \etal 1991b} {\PR}{D44}{3272}
\refjl{Gupta R \etal 1992} {preprint LA-UR-91-3522}{}{}
\refjl{Gupta S \etal 1990}{\PL}{B242}{437}
\refjl{Gupta S 1992a}{\NP}{B370}{741}
\refjl{Gupta S 1992b}{\PL}{B288}{171}

\refjl{Hands S J \etal 1992}{preprint ILL-TH-92-16}{}{}
\refjl{Hasenbusch M 1990}{\NP}{B333}{581}
\refjl{Hasenfratz A and Hasenfratz P 1986}{\PR}{D34}{3160}
\refjl{Hasenfratz A \etal 1987}{\PL}{B199}{531}
\refjl{Hasenfratz A and Neuhaus T 1988}{\NP}{B297}{205}
\refjl{Hasenfratz A \etal 1990}{\ZP}{C46}{257}
\refjl{Hasenfratz P and Leutwyler H 1990}{\NP}{B343}{241}
\refjl{Hashimoto S and Saeki Y 1992}{\NP (Proc. Suppl.)}{B26}{381}
\refbk{Herrmann H J and Karsch F (eds) 1991}
      {Fermion Algorithms}{World Scientific, Singapore}
\refjl{Horowitz A M 1990}{\PL}{B244}{306}
\refjl{Horsley R \etal 1989}{\NP}{B313}{377}
\refjl{Huntley A and Michael C 1987}{\NP}{B286}{211}

\refjl{Iwasaki Y \etal (QCDPAX Collaboration) 1991}{\PRL}{67}{3343}
\refjl{Iwasaki Y \etal (QCDPAX Collaboration) 1992a}
      {preprint UTHEP-246}{}{}
\refjl{Iwasaki Y \etal 1992b}{preprint UTHEP-237}{}{}

\refjl{Jansen K and Seuferling P 1990}{\NP}{B343}{507}

\refjl{Kalkreuter T 1992}{preprint DESY 92-158}{}{}
\refjl{Kaplan D B 1992}{\PL}{B288}{342}
\refjl{Kapusta J I 1979}{\NP}{B148}{461}
\refjl{Karsch F \etal 1987}{\PL}{B188}{353}
\refjl{Karsch F \etal 1988}{\ZP}{C37}{617}
\refjl{Karsch F and Wyld H W 1988}{\PL}{B213}{505}
\refjl{Kawai H 1992}{\NP (Proc. Suppl.)}{B26}{93}
\refjl{Kennedy A D and Pendleton B J 1990}{\NP (Proc. Suppl.)}{B20}{118}
\refjl{Kilcup G 1991}{\NP (Proc. Suppl.)}{B20}{417}
\refjl{Kirzhnits D A and Linde A D 1972}{\PL}{B42}{471}
\refjl{Kocic A \etal 1992a}{preprint CERN-TH 6542/92}{}{}
\refjl{Kocic A \etal 1992b}{\PL}{B289}{400}
\refjl{Kocic A \etal 1992c}{preprint ILL-TH-92-17}{}{}
\refjl{Kodama K \etal (E653 Collaboration) 1992}{\PL}{B286}{187}
\refjl{Kogut J B and Susskind L 1975}{\PR}{D11}{395}
\refjl{Kogut J B \etal 1988}{\PRL}{60}{772}
\refjl{Kogut J B \etal 1989}{\NP}{B317}{253, 271}
\refjl{Kogut J B \etal 1991}{\PR}{D44}{2869}
\refjl{Koller J and van Baal P 1988}{\NP}{B302}{1}
\refbk{Kondo K-I 1991 in}{1990 International Workshop on
       Strong Coupling Gauge Theories and Beyond}{Muta T and
       Yamawaki K (eds), World Scientific, Singapore}
\refbk{Kripfganz J 1992 in}{Dynamics of First Order Phase
       Transitions}{Herrmann H J, Janke W and Karsch F (eds)
       World Scientific, Singapore}
\refjl{K\"uhn J H and Zerwas P M 1988}{Phys. Rep.}{167}{321}
\refjl{Kuti J \etal 1988}{\PRL}{61}{678}

\refjl{Lacock P \etal 1992}{\NP}{B369}{501}
\refjl{Laermann E \etal ($MT_c$ Collaboration) 1992}
       {\NP (Proc. Suppl.)}{B26}{268}
\refjl{Landsman N P 1988}{\NP}{B322}{498}
\refjl{Langhammer F 1986}{Diplomarbeit RWTH Aachen}{}{}
\refjl{Langguth W and Montvay I 1987}{\ZP}{C36}{725}
\refjl{Lee B W \etal 1977}{\PR}{D16}{1519}
\refjl{Lepage P and Mackenzie P 1992}{Preprint FERMI-91/355-T}{}{}
\refjl{Lin L \etal 1990}{\ZP}{C48}{355}
\refjl{Lin L \etal 1991a}{\PL}{B264}{407}
\refjl{Lin L \etal 1991b}{\NP}{B355}{511}
\refjl{Linde A D 1980}{\PL}{B96}{289}
\refjl{Lombardo M-P \etal 1992}{preprint ILL-TH-92-15}{}{}
\refjl{L\"uscher M \etal 1980}{\NP}{B173}{365}
\refjl{L\"uscher M and Weisz P 1987}{\NP}{B290}{25}
\refjl{L\"uscher M and Weisz P 1988}{\NP}{B295}{65}
\refjl{L\"uscher M and Weisz P 1989}{\NP}{B318}{705}
\refjl{L\"uscher M \etal 1993}{\NP}{B389}{247}
\refjl{Lubicz V \etal 1991} {\NP}{B356}{301}
\refjl{Lubicz V \etal 1992} {\PL}{B274}{415}

\refjl{Maiani L and Testa M 1990}{\PL}{B245}{585}
\refjl{Manousakis E and Polonyi J 1987}{\PRL}{58}{847}
\refbk{Marinari E 1992 in}{Proceedings of the International Symposium
       on Lattice Field Theory}{Amsterdam}
\refjl{Martinelli G \etal 1991}{\NP}{B358}{211}
\refjl{Martinelli G \etal 1992}{\NP}{B378}{591}
\refjl{Matsui T and Satz H 1986}{\PL}{B178}{416}
\refbk{Mawhinney R 1992 in}{Proceedings of the International Symposium
       on Lattice Field Theory}{Amsterdam}
\refjl{Metropolis N \etal 1953}{\JCP}{21}{1087}
\refjl{Michael C 1992}{\PL}{B283}{103}
\refjl{Michael C and Teper M 1989}{\NP}{B314}{347}
\refjl{Miransky V A 1985}{\NC}{90A}{149}
\refjl{Montvay I 1987a}{\PL}{B199}{89}
\refjl{Montvay I 1987b}{\PL}{B205}{315}
\refjl{Montvay I 1992}{\NP (Proc. Suppl.)}{B26}{57}

\refbk{Negele J 1992 in}{Proceedings of the International Symposium
       on Lattice Field Theory}{Amsterdam}
\refjl{Neuberger H 1988}{\NP}{B300}{180}
\refjl{Nielsen H R and Ninomiya M 1981}{\NP}{B193}{173}

\refbk{Parisi G 1980 in}{Proceedings of the XXth Conference
       on High Energy Physics}{Madison}
\refjl{Parisi G 1992}{\NP (Proc. Suppl.)}{B26}{181}
\refjl{Parisi G and Wu Y S 1981}{Sci. Sin}{24}{483}
\refjl{Patrascioiu A and Seiler E 1992}{preprint MPI-Ph/92-18,
       LPTHE 92/63}{}{}
\refjl{Petcher D N and Weingarten D H 1981}{\PL}{B99}{333}
\refjl{Petersson B 1992}{preprint BI-TP-92/58}{}{}
\refjl{Pisarski R D and Wilczek F 1984}{\PR}{D29}{338}
\refjl{Polyakov A M 1979}{\PL}{B82}{2410}

\refjl{Rakow P E L 1991}{\NP}{B356}{27}
\refjl{Rakow P E L 1992}{preprint FUB-HEP 22/92}{}{}
\refbk{Rebbi C (ed) 1983}{Lattice Gauge Theories and Monte Carlo
       Simulations}{World Scientific, Singapore}
\refjl{Reisz T 1991}{\JMP}{32}{115}
\refjl{Reisz T 1992}{\ZP}{53}{169}

\refjl{Scalettar R T \etal 1986}{\PR}{B34}{7911}
\refjl{Sexton J C and Weingarten D H 1992}{\NP}{B380}{665}
\refjl{Shaposhnikov M 1992}{\NP (Proc. Suppl.)}{B26}{78}
\refjl{Sharatchandra H S  \etal 1981}{\NP}{B192}{205}
\refjl{Sharpe S \etal 1992a}{\NP}{B383}{309}
\refjl{Sharpe S \etal 1992b}{\NP (Proc. Suppl.)}{B26}{197}
\refjl{Sheikoleslami B and Wohlert R 1985}{\NP}{B259}{572}
\refjl{Smit J 1980}{\NP}{B175}{307}
\refjl{Smit J 1986}{Acta Phys. Polon.}{B17}{531}
\refjl{Smit J 1989}{\NP (Proc. Suppl.)}{B9}{579}
\refjl{Svetitsky B and Yaffe L G 1982a}{\PR}{D26}{963}
\refjl{Svetitsky B and Yaffe L G 1982b}{\NP}{B210}{423}
\refjl{Swendsen R H and Wang J-S 1987}{\PRL}{58}{86}
\refjl{Swift P V D 1984}{\PL}{B145}{256}
\refjl{Symanzik K 1983a} {\NP}{B226}{187}
\refjl{Symanzik K 1983b} {\NP}{B226}{205}

\refjl{Teper M 1986}{\PL}{B183}{345}
\refjl{Thornton A M 1989}{\PL}{B221}{151}

\refjl{Unger L 1992}{preprint CU-TP-571}{}{}

\refjl{van Baal P and Kronfeld A 1989}{\NP (Proc. Suppl.)}{B9}{227}

\refjl{Vohwinkel C 1989}{\PRL}{63}{2544}

\refjl{Weinberg S 1974}{\PR}{D9}{3320}
\refjl{Weisz P 1981}{\PL}{100B}{331}
\refjl{Wilczek F 1992}{Int. J. Mod. Phys.}{A7}{3911}
\refjl{Wilson K 1974}{\PR}{D10}{2445}
\refjl{Wolff U 1989}{\NP}{B322}{759}

\figures

\figcaption{$T_c/\sqrt{\sigma}$ for $SU(2)$ (circles)
and $SU(3)$ (triangles) pure gauge theory
as well as 4-flavour QCD (cross).
}

\figcaption{$T_c$ in MeV extracted from
$m_{\rho}$ for QCD with $n_f$=0 (dots),
2 (triangles) and 4 (square) flavours of light quarks.
}

\figcaption{$T_c/\Lambda_{\overline{\rm MS}}$ (a) and $\sqrt{\sigma} /
\Lambda_{\overline{\rm MS}}$ (b) for the
$SU(3)$ pure gauge theory obtained by using \eq{rge} with
the bare coupling, $\beta$ (triangles), and the effective coupling
$\beta_E$ (circles).
}

\figcaption{The heavy quark potential in quenched QCD (data obtained
from Bali and Schilling 1992).
}

\figcaption{The spin dependent potentials in full QCD, taken
from (Laermann \etal~1992). The results were obtained with a quark
mass of $m_q\simeq 35$ MeV. The dotted lines, except for $V_1^\prime
(R)$, indicate the perturbative prediction of a short-ranged
vector potential.}

\figcaption{The pion mass squared as a function of the quark
mass $m$ (boxes). The second data set (circles) are the results
for one of the additional pseudo scalar mesons in the staggered
fermion discretization which also become Goldstone bosons in
the continuum limit.
}

\figcaption{The Edinburgh plot for quenched Wilson fermions.
The filled diamonds denote the static and the experimental result
respectively.
}

\figcaption{Colour $SU(2)$ data for glueball masses.
$T_2^+$ and $E_2^+$ denote two different lattice representations
embedded in the continuum $2^+$ representation. The masses are
normalized to the $0^+$ glueball $(A_1^+)$ mass. Also shown is the
string tension $\sqrt \kappa$ and the torelon $(T_{11}^+)$ mass. The
solid lines are the analytic predictions. The plot is taken from
(van Baal and Kronfeld 1989).
}

\figcaption{$f_{PS} \sqrt{M_{PS}}$ as function of the inverse
heavy-light pseudoscalar mass, $1/M_{PS}$. The results of several
calculations at various $\beta$ with propagating quarks and in the
static limit are shown. The curve denotes the result of a
quadratic fit to the corrections of the scaling law
$f_{PS}\sqrt{M_{PS}} =$ const.
}

\figcaption{
In (a) energy density and pressure of an $SU(3)$ gauge theory in
units of $T^4$ are shown
as a function of $T/T_c$. The curves have been obtained from an integration
of data for the free energy density (Engels \etal~1990a). Violations of
asymptotic scaling have been taken into account. In (b) the same
quantities are shown for two flavour QCD (Gottlieb \etal~1987d). So far
no attempt has been made to correct for scaling violations in this case.}

\figcaption{
Determination of the critical exponent $\nu$ for the $SU(2)$
deconfinement transition obtained from a finite size scaling
analysis of the order parameter (Engels \etal~1990b) (a). In
(b) the finite size scaling of the critical coupling is shown.
This leads to a determination of the ratio of critical exponents,
$\gamma / \nu$, for the $SU(3)$ gauge theory (Fukugita \etal~1990).}

\figcaption{
Screening masses in units of the lattice spacing
for the pure $SU(3)$ gauge theory obtained
on lattices of size $4 \times 24^3$ (squares)  (Fukugita \etal~1989)
(dots) (Brown \etal~1988) and
$4 \times 8^2 \times 32$ (triangles) (Bacilieri \etal~1988).}

\figcaption{
Generic phase diagram for three flavour QCD (Brown \etal~1990).
The arcs enclose regions of
first order phase transitions. The dashed lines indicate lines of
2$^{\rm nd}$ order phase transitions.
On the m$_u$=0 axis this has been displaced a bit
for better visibility.
The crossed circle indicates the
location of the physical strange to up quark mass ratio. Present Monte
Carlo simulations suggest that this point lies outside the first order
region. However, this still has to be verified on larger lattices.}

\figcaption{
Screening masses $\mu${\its /T} as a function of $\beta$ for $n_f$=4 from
an 8$\times$16$^3$ lattice. The state denoted ``q" gives the screening
mass extracted from the quark propagator in Landau gauge
(Boyd \etal~1992b).
Lines to the left give the values of the zero temperature masses in
units of {\its T}$_c$ calculated at $\beta$=5.15;
{\its m}$_{\pi}${\its /T}$_c$ (dotted),
{\its m}$_{\sigma}${\its /T}$_c$ (dashed-dotted),
{\its m}$_{\rho}${\its /T}$_c$ (short dashes),
and {\its m}$_{N}${\its /T}$_c$ (long dashes).
Lines to the right give screening masses corresponding to free quark
propagation in the quark (solid),
mesonic (short dashes) and baryonic (long dashes) channels.}

\figcaption{Upper bound on the ratio $m_H / \langle \phi
\rangle$ in the pure Higgs sector.
The data are obtained with the standard action (diamonds)
(Hasenfratz A \etal 1987, G\"ockeler \etal 1992a)
and on simplicial lattices (circles)(Bhanot \etal 1990).
The curve denotes the
analytic result (L\"uscher and Weisz 1989)
for the standard action with the estimated
theoretical uncertainty indicated by error bars.
As the cut-off definition, $\Lambda = 1/a$ has been adopted.}

\end
%
%

\font\titlefont=cmbx10 scaled \magstep1

\font\sectnfont=cmbx8  scaled \magstep1
\def\mname{\ifcase\month\or January \or February \or March \or April
           \or May \or June \or July \or August \or September
           \or October \or November \or December \fi}
\def\date{\hbox{\strut\mname \number\year}}
\def\banner{\hfill\hbox{\vbox{\offinterlineskip\crnum}}\relax}
\def\manner{\hbox{\vbox{\offinterlineskip\crnum\date}}
               \hfill\relax}
\footline={\ifnum
\pageno=0
\ \else\hfil\number\pageno\hfil\fi}
%
%
%
\newcount\SECTIONNUMBER\SECTIONNUMBER=0
\def\section#1{\global\advance\SECTIONNUMBER by 1\SUBSECTIONNUMBER=0
      \bigskip\goodbreak\line{{\sectnfont \the\SECTIONNUMBER.\ #1}\hfil}
      \smallskip}

\newcount\FIGURENUMBER\FIGURENUMBER=0
\def\FIG#1{\expandafter\ifx\csname FG#1\endcsname\relax
               \global\advance\FIGURENUMBER by 1
               \expandafter\xdef\csname FG#1\endcsname
                              {\the\FIGURENUMBER}\fi}
\def\figtag#1{\expandafter\ifx\csname FG#1\endcsname\relax
               \global\advance\FIGURENUMBER by 1
               \expandafter\xdef\csname FG#1\endcsname
                              {\the\FIGURENUMBER}\fi
              \csname FG#1\endcsname\relax}
\def\fig#1{\expandafter\ifx\csname FG#1\endcsname\relax
               \global\advance\FIGURENUMBER by 1
               \expandafter\xdef\csname FG#1\endcsname
                      {\the\FIGURENUMBER}\fi
           figure~\csname FG#1\endcsname\relax}
\def\Fig#1{\expandafter\ifx\csname FG#1\endcsname\relax
               \global\advance\FIGURENUMBER by 1
               \expandafter\xdef\csname FG#1\endcsname
                      {\the\FIGURENUMBER}\fi
           Figure~\csname FG#1\endcsname\relax}
\def\figand#1#2{\expandafter\ifx\csname FG#1\endcsname\relax
               \global\advance\FIGURENUMBER by 1
               \expandafter\xdef\csname FG#1\endcsname
                      {\the\FIGURENUMBER}\fi
           \expandafter\ifx\csname FG#2\endcsname\relax
               \global\advance\FIGURENUMBER by 1
               \expandafter\xdef\csname FG#2\endcsname
                      {\the\FIGURENUMBER}\fi
           figures \csname FG#1\endcsname\ and
                   \csname FG#2\endcsname\relax}
\def\figto#1#2{\expandafter\ifx\csname FG#1\endcsname\relax
               \global\advance\FIGURENUMBER by 1
               \expandafter\xdef\csname FG#1\endcsname
                      {\the\FIGURENUMBER}\fi
           \expandafter\ifx\csname FG#2\endcsname\relax
               \global\advance\FIGURENUMBER by 1
               \expandafter\xdef\csname FG#2\endcsname
                      {\the\FIGURENUMBER}\fi
           figures \csname FG#1\endcsname--\csname FG#2\endcsname\relax}
\newcount\TABLENUMBER\TABLENUMBER=0
\def\TABLE#1{\expandafter\ifx\csname TB#1\endcsname\relax
               \global\advance\TABLENUMBER by 1
               \expandafter\xdef\csname TB#1\endcsname
                          {\the\TABLENUMBER}\fi}
\def\tabletag#1{\expandafter\ifx\csname TB#1\endcsname\relax
               \global\advance\TABLENUMBER by 1
               \expandafter\xdef\csname TB#1\endcsname
                          {\the\TABLENUMBER}\fi
             \csname TB#1\endcsname\relax}
\def\table#1{\expandafter\ifx\csname TB#1\endcsname\relax
               \global\advance\TABLENUMBER by 1
               \expandafter\xdef\csname TB#1\endcsname{\the\TABLENUMBER}\fi
              \csname TB#1\endcsname\relax}
\def\tableand#1#2{\expandafter\ifx\csname TB#1\endcsname\relax
               \global\advance\TABLENUMBER by 1
               \expandafter\xdef\csname TB#1\endcsname{\the\TABLENUMBER}\fi
             \expandafter\ifx\csname TB#2\endcsname\relax
               \global\advance\TABLENUMBER by 1
               \expandafter\xdef\csname TB#2\endcsname{\the\TABLENUMBER}\fi
             tables \csname TB#1\endcsname{} and
                    \csname TB#2\endcsname\relax}
\def\tableto#1#2{\expandafter\ifx\csname TB#1\endcsname\relax
               \global\advance\TABLENUMBER by 1
               \expandafter\xdef\csname TB#1\endcsname{\the\TABLENUMBER}\fi
             \expandafter\ifx\csname TB#2\endcsname\relax
               \global\advance\TABLENUMBER by 1
               \expandafter\xdef\csname TB#2\endcsname{\the\TABLENUMBER}\fi
            tables \csname TB#1\endcsname--\csname TB#2\endcsname\relax}
\newcount\REFERENCENUMBER\REFERENCENUMBER=0
\def\REF#1{\expandafter\ifx\csname RF#1\endcsname\relax
               \global\advance\REFERENCENUMBER by 1
               \expandafter\xdef\csname RF#1\endcsname
                         {\the\REFERENCENUMBER}\fi}
\def\reftag#1{\expandafter\ifx\csname RF#1\endcsname\relax
               \global\advance\REFERENCENUMBER by 1
               \expandafter\xdef\csname RF#1\endcsname
                      {\the\REFERENCENUMBER}\fi
             \csname RF#1\endcsname\relax}
\def\ref#1{\expandafter\ifx\csname RF#1\endcsname\relax
               \global\advance\REFERENCENUMBER by 1
               \expandafter\xdef\csname RF#1\endcsname
                      {\the\REFERENCENUMBER}\fi
             [\csname RF#1\endcsname]\relax}
\def\refto#1#2{\expandafter\ifx\csname RF#1\endcsname\relax
               \global\advance\REFERENCENUMBER by 1
               \expandafter\xdef\csname RF#1\endcsname
                      {\the\REFERENCENUMBER}\fi
               \expandafter\ifx\csname RF#2\endcsname\relax
               \global\advance\REFERENCENUMBER by 1
               \expandafter\xdef\csname RF#2\endcsname
                      {\the\REFERENCENUMBER}\fi
             [\csname RF#1\endcsname--\csname RF#2\endcsname]\relax}
\def\refand#1#2{\expandafter\ifx\csname RF#1\endcsname\relax
               \global\advance\REFERENCENUMBER by 1
               \expandafter\xdef\csname RF#1\endcsname
                      {\the\REFERENCENUMBER}\fi
           \expandafter\ifx\csname RF#2\endcsname\relax
               \global\advance\REFERENCENUMBER by 1
               \expandafter\xdef\csname RF#2\endcsname
                      {\the\REFERENCENUMBER}\fi
            [\csname RF#1\endcsname,\csname RF#2\endcsname]\relax}
\def\refs#1#2{\expandafter\ifx\csname RF#1\endcsname\relax
               \global\advance\REFERENCENUMBER by 1
               \expandafter\xdef\csname RF#1\endcsname
                      {\the\REFERENCENUMBER}\fi
           \expandafter\ifx\csname RF#2\endcsname\relax
               \global\advance\REFERENCENUMBER by 1
               \expandafter\xdef\csname RF#2\endcsname
                      {\the\REFERENCENUMBER}\fi
            [\csname RF#1\endcsname,\csname RF#2\endcsname]\relax}
\def\refss#1#2#3{\expandafter\ifx\csname RF#1\endcsname\relax
               \global\advance\REFERENCENUMBER by 1
               \expandafter\xdef\csname RF#1\endcsname
                      {\the\REFERENCENUMBER}\fi
           \expandafter\ifx\csname RF#2\endcsname\relax
               \global\advance\REFERENCENUMBER by 1
               \expandafter\xdef\csname RF#2\endcsname
                      {\the\REFERENCENUMBER}\fi
           \expandafter\ifx\csname RF#3\endcsname\relax
               \global\advance\REFERENCENUMBER by 1
               \expandafter\xdef\csname RF#3\endcsname
                      {\the\REFERENCENUMBER}\fi
     [\csname RF#1\endcsname,\csname RF#2\endcsname,\csname
              RF#3\endcsname]\relax}
\def\Ref#1{\expandafter\ifx\csname RF#1\endcsname\relax
               \global\advance\REFERENCENUMBER by 1
               \expandafter\xdef\csname RF#1\endcsname
                      {\the\REFERENCENUMBER}\fi
             Ref.~\csname RF#1\endcsname\relax}
\def\Refs#1#2{\expandafter\ifx\csname RF#1\endcsname\relax
               \global\advance\REFERENCENUMBER by 1
               \expandafter\xdef\csname RF#1\endcsname
                      {\the\REFERENCENUMBER}\fi
           \expandafter\ifx\csname RF#2\endcsname\relax
               \global\advance\REFERENCENUMBER by 1
               \expandafter\xdef\csname RF#2\endcsname
                      {\the\REFERENCENUMBER}\fi
        Refs.~\csname RF#1\endcsname{},\csname RF#2\endcsname\relax}
\def\Refto#1#2{\expandafter\ifx\csname RF#1\endcsname\relax
               \global\advance\REFERENCENUMBER by 1
               \expandafter\xdef\csname RF#1\endcsname
                      {\the\REFERENCENUMBER}\fi
           \expandafter\ifx\csname RF#2\endcsname\relax
               \global\advance\REFERENCENUMBER by 1
               \expandafter\xdef\csname RF#2\endcsname
                      {\the\REFERENCENUMBER}\fi
            Refs.~\csname RF#1\endcsname--\csname RF#2\endcsname]\relax}
\def\Refand#1#2{\expandafter\ifx\csname RF#1\endcsname\relax
               \global\advance\REFERENCENUMBER by 1
               \expandafter\xdef\csname RF#1\endcsname
                      {\the\REFERENCENUMBER}\fi
           \expandafter\ifx\csname RF#2\endcsname\relax
               \global\advance\REFERENCENUMBER by 1
               \expandafter\xdef\csname RF#2\endcsname
                      {\the\REFERENCENUMBER}\fi
        Refs.~\csname RF#1\endcsname{} and \csname RF#2\endcsname\relax}
\newcount\EQUATIONNUMBER\EQUATIONNUMBER=0
\def\EQ#1{\expandafter\ifx\csname EQ#1\endcsname\relax
               \global\advance\EQUATIONNUMBER by 1
               \expandafter\xdef\csname EQ#1\endcsname
                          {\the\EQUATIONNUMBER}\fi}
\def\eqtag#1{\expandafter\ifx\csname EQ#1\endcsname\relax
               \global\advance\EQUATIONNUMBER by 1
               \expandafter\xdef\csname EQ#1\endcsname
                      {\the\EQUATIONNUMBER}\fi
            \csname EQ#1\endcsname\relax}
\def\EQNO#1{\expandafter\ifx\csname EQ#1\endcsname\relax
            \global\advance\EQUATIONNUMBER by 1
            \expandafter\xdef\csname EQ#1\endcsname
                      {\the\secno.\the\EQUATIONNUMBER}\fi
            \eqno(\csname EQ#1\endcsname)\relax}
\def\EQNM#1{\expandafter\ifx\csname EQ#1\endcsname\relax
               \global\advance\EQUATIONNUMBER by 1
               \expandafter\xdef\csname EQ#1\endcsname
                      {\the\EQUATIONNUMBER}\fi
            (\csname EQ#1\endcsname)\relax}
\def\eq#1{\expandafter\ifx\csname EQ#1\endcsname\relax
               \global\advance\EQUATIONNUMBER by 1
               \expandafter\xdef\csname EQ#1\endcsname
                      {\the\EQUATIONNUMBER}\fi
          (\csname EQ#1\endcsname)\relax}
\def\eqand#1#2{\expandafter\ifx\csname EQ#1\endcsname\relax
               \global\advance\EQUATIONNUMBER by 1
               \expandafter\xdef\csname EQ#1\endcsname
                        {\the\EQUATIONNUMBER}\fi
          \expandafter\ifx\csname EQ#2\endcsname\relax
               \global\advance\EQUATIONNUMBER by 1
               \expandafter\xdef\csname EQ#2\endcsname
                      {\the\EQUATIONNUMBER}\fi
         \csname EQ#1\endcsname{} and \csname EQ#2\endcsname\relax}
\def\eqto#1#2{\expandafter\ifx\csname EQ#1\endcsname\relax
               \global\advance\EQUATIONNUMBER by 1
               \expandafter\xdef\csname EQ#1\endcsname
                      {\the\EQUATIONNUMBER}\fi
          \expandafter\ifx\csname EQ#2\endcsname\relax
               \global\advance\EQUATIONNUMBER by 1
               \expandafter\xdef\csname EQ#2\endcsname
                      {\the\EQUATIONNUMBER}\fi
          \csname EQ#1\endcsname--\csname EQ#2\endcsname\relax}
%
\newcount\SUBSECTIONNUMBER\SUBSECTIONNUMBER=0
\def\subsection#1{\global\advance\SUBSECTIONNUMBER by 1
      \bigskip\goodbreak\line{{\sectnfont
         \the\SECTIONNUMBER.\the\SUBSECTIONNUMBER.\ #1}\hfil}
      \smallskip}
%

\def\NP{{\sl Nucl.\ Phys.\ }}
\def\PL{{\sl Phys.\ Lett.\ }}
\def\PR{{\sl Phys.\ Rev.\ }}

\def\PRL{{\sl Phys.\ Rev.\ Lett.\ }}

\def\ZP{{\sl Z.\ Phys.\ }}
%
%
\def\DT{\Delta\tau}\def\dt{\ifmmode\DT\else$\DT$\fi}
\def\BETAC{\beta_c}\def\betac{\ifmmode\BETAC\else$\BETAC$\fi}
\def\LARGE{8\times16^3}\def\large{\ifmmode\LARGE\else$\LARGE$\fi}
\def\SMALL{8\times12^3}\def\small{\ifmmode\SMALL\else$\SMALL$\fi}
\def\MQ{ma}\def\mq{\ifmmode\MQ\else$\MQ$\fi}
\def\REL{Re(L)}\def\rel{\ifmmode\REL\else$\REL$\fi}
\def\ABL{Abs(L)}\def\absl{\ifmmode\ABL\else$\ABL$\fi}
\def\PPBAR{\overline\chi\chi}\def\ppbar{\ifmmode\PPBAR\else$\PPBAR$\fi}
\def\ord{\ifmmode\langle\REL\rangle\else$\langle\REL\rangle$\fi}
\def\cord{\ifmmode\langle\PPBAR\rangle\else$\langle\PPBAR\rangle$\fi}
\def\pprgi{\ifmmode\langle\PPBAR\rangle_{RGI}\else$\langle\PPBAR
           \rangle_{RGI}$\fi}
\def\entropy{\ifmmode{s\over T^3}
             \else$s/T^3$\fi}
\def\NT{N_\tau}\def\nt{\ifmmode\NT\else$\NT$\fi}
\def\NS{N_\sigma}\def\ns{\ifmmode\NS\else$\NS$\fi}
\def\GT{g_\tau}\def\gt{\ifmmode\GT\else$\GT$\fi}
\def\GS{g_\sigma}\def\gs{\ifmmode\GS\else$\GS$\fi}
\def\CT{c'_\tau}\def\ct{\ifmmode\CT\else$\CT$\fi}
\def\CS{c'_\sigma}\def\cs{\ifmmode\CS\else$\CS$\fi}
\def\NF{n_\f}\def\nf{\ifmmode\NF\else$\NF$\fi}
\def\TC{T_c}\def\tc{\ifmmode\TC\else$\TC$\fi}
\def\bar{\overline}

\def\lsim{\raise0.3ex\hbox{$<$\kern-0.75em\raise-1.1ex\hbox{$\sim$}}}
\def\gsim{\raise0.3ex\hbox{$>$\kern-0.75em\raise-1.1ex\hbox{$\sim$}}}
%

\def\f{{\scriptscriptstyle f}}
%
\def\ie{{i.~e.\/}}

%
%
%
%
%
%
%
%
%


\magnification=1200
\hsize=31pc
\vsize=55 truepc
\hfuzz=2pt
\vfuzz=4pt
\pretolerance=5000
\tolerance=5000
\parskip=0pt plus 1pt
\parindent=16pt
%

%
%
\font\fourteenrm=cmr10 scaled \magstep2
\font\fourteeni=cmmi10 scaled \magstep2
\font\fourteenbf=cmbx10 scaled \magstep2
\font\fourteenit=cmti10 scaled \magstep2
\font\fourteensy=cmsy10 scaled \magstep2

%
\font\large=cmbx10 scaled \magstep1

%

%

%

%
\font\eightrm=cmr8
\font\eighti=cmmi8
\font\eightbf=cmbx8
\font\eightit=cmti8

\font\eightsy=cmsy8
\font\sixrm=cmr6
\font\sixi=cmmi6
\font\sixsy=cmsy6

%
\def\tenpoint{\def\rm{\fam0\tenrm}%
  \textfont0=\tenrm \scriptfont0=\sevenrm
                      \scriptscriptfont0=\fiverm
  \textfont1=\teni  \scriptfont1=\seveni
                      \scriptscriptfont1=\fivei
  \textfont2=\tensy \scriptfont2=\sevensy
                      \scriptscriptfont2=\fivesy
  \textfont3=\tenex   \scriptfont3=\tenex
                      \scriptscriptfont3=\tenex
  \textfont\itfam=\tenit  \def\it{\fam\itfam\tenit}%
  \textfont\slfam=\tensl  \def\sl{\fam\slfam\tensl}%
  \textfont\bffam=\tenbf  \scriptfont\bffam=\sevenbf
                            \scriptscriptfont\bffam=\fivebf
                            \def\bf{\fam\bffam\tenbf}%
  \normalbaselineskip=20 truept
  \setbox\strutbox=\hbox{\vrule height14pt depth6pt
width0pt}%
  \let\sc=\eightrm \normalbaselines\rm}
\def\eightpoint{\def\rm{\fam0\eightrm}%
  \textfont0=\eightrm \scriptfont0=\sixrm
                      \scriptscriptfont0=\fiverm
  \textfont1=\eighti  \scriptfont1=\sixi
                      \scriptscriptfont1=\fivei
  \textfont2=\eightsy \scriptfont2=\sixsy
                      \scriptscriptfont2=\fivesy
  \textfont3=\tenex   \scriptfont3=\tenex
                      \scriptscriptfont3=\tenex
  \textfont\itfam=\eightit  \def\it{\fam\itfam\eightit}%
  \textfont\bffam=\eightbf  \def\bf{\fam\bffam\eightbf}%
  \normalbaselineskip=16 truept
  \setbox\strutbox=\hbox{\vrule height11pt depth5pt width0pt}}
\def\fourteenpoint{\def\rm{\fam0\fourteenrm}%
  \textfont0=\fourteenrm \scriptfont0=\tenrm
                      \scriptscriptfont0=\eightrm
  \textfont1=\fourteeni  \scriptfont1=\teni
                      \scriptscriptfont1=\eighti
  \textfont2=\fourteensy \scriptfont2=\tensy
                      \scriptscriptfont2=\eightsy
  \textfont3=\tenex   \scriptfont3=\tenex
                      \scriptscriptfont3=\tenex
  \textfont\itfam=\fourteenit  \def\it{\fam\itfam\fourteenit}%
  \textfont\bffam=\fourteenbf  \scriptfont\bffam=\tenbf
                             \scriptscriptfont\bffam=\eightbf
                             \def\bf{\fam\bffam\fourteenbf}%
  \normalbaselineskip=24 truept
  \setbox\strutbox=\hbox{\vrule height17pt depth7pt width0pt}%
  \let\sc=\tenrm \normalbaselines\rm}
\def\today{\number\day\ \ifcase\month\or
  January\or February\or March\or April\or May\or June\or
  July\or August\or September\or October\or November\or
December\fi
  \space \number\year}

%
\newcount\secno      
\newcount\subno      
\newcount\subsubno   
\newcount\appno      
\newcount\tableno    
\newcount\figureno   
%

%
\normalbaselineskip=20 truept
\baselineskip=20 truept

%
%
\def\title#1
   {\vglue1truein
   {\baselineskip=24 truept
    \pretolerance=10000
    \raggedright
    \noindent \fourteenpoint\bf #1\par}
    \vskip1truein minus36pt}
%

%
\def\author#1
  {{\pretolerance=10000
    \raggedright
    \noindent {\large #1}\par}}

%
\def\address#1
   {\bigskip
    \noindent \rm #1\par}

%
\def\shorttitle#1
   {\vfill
    \noindent \rm Short title: {\sl #1}\par
    \medskip}

%
\def\pacs#1
   {\noindent \rm PACS number(s): #1\par
    \medskip}

%
\def\jnl#1
   {\noindent \rm Submitted to: {\sl #1}\par
    \medskip}

%
\def\date
   {\noindent Date: \today\par
    \medskip}

%

%
\def\keyword#1
   {\bigskip
    \noindent {\bf Keyword abstract: }\rm#1}

%

%
%
\def\contents
   {{\noindent
    \bf Contents
    \par}
    \rightline{Page}}

%
\def\entry#1#2#3
   {\noindent
    \hangindent=20pt
    \hangafter=1
    \hbox to20pt{#1 \hss}#2\hfill #3\par}

%
\def\subentry#1#2#3
   {\noindent
    \hangindent=40pt
    \hangafter=1
    \hskip20pt\hbox to20pt{#1 \hss}#2\hfill #3\par}
\def\checkforsub{\futurelet\nexttok\decide}
\def\ssf{\relax}
\def\decide{\if\nexttok\ssf\let\endspace=\nospace
                \else\let\endspace=\extraspace\fi\endspace}
\def\nospace{\nobreak\par\nobreak}
%
%
\def\section#1{%
    \goodbreak
    \vskip50pt plus12pt minus12pt
    \nobreak
    \gdef\extraspace{\nobreak\bigskip\noindent\ignorespaces}%
    \noindent
    \subno=0 \subsubno=0 \EQUATIONNUMBER=0
    \global\advance\secno by 1
    \noindent {\bf \the\secno. #1}\par\checkforsub}

%
\def\subsection#1{%
     \goodbreak
     \vskip24pt plus12pt minus6pt
     \nobreak
     \gdef\extraspace{\nobreak\medskip\noindent\ignorespaces}%
     \noindent
     \subsubno=0
     \global\advance\subno by 1
     \noindent {\sl \the\secno.\the\subno. #1\par}\checkforsub}

%
\def\subsubsection#1{%
     \goodbreak
     \vskip20pt plus6pt minus6pt
     \nobreak\noindent
     \global\advance\subsubno by 1
     \noindent {\sl \the\secno.\the\subno.\the\subsubno. #1}\null.
     \ignorespaces}

%
\def\appendix#1
   {\vskip0pt plus.1\vsize\penalty-250
    \vskip0pt plus-.1\vsize\vskip24pt plus12pt minus6pt
    \subno=0
    \global\advance\appno by 1
    \noindent {\bf Appendix \the\appno. #1\par}
    \bigskip
    \noindent}

%
\def\subappendix#1
   {\vskip-\lastskip
    \vskip36pt plus12pt minus12pt
    \bigbreak
    \global\advance\subno by 1
    \noindent {\sl \the\appno.\the\subno. #1\par}
    \nobreak
    \medskip
    \noindent}

%
\def\ack
   {\vskip-\lastskip
    \vskip36pt plus12pt minus12pt
    \bigbreak
    \noindent{\bf Acknowledgments\par}
    \nobreak
    \bigskip
    \noindent}


%

%
\def\tabcaption#1
   {\global\advance\tableno by 1
    \noindent {\bf Table \the\tableno.} \rm#1\par
    \bigskip}

%

%

%
\def\ns{\noalign{\vskip-6pt}}

%

%
\def\figures
   {\vfill\eject
    \noindent {\bf Figure captions\par}
    \bigskip}

%
\def\figcaption#1
   {\global\advance\figureno by 1
    \noindent {\bf Figure \the\figureno.} \rm#1\par
    \bigskip}

%
\def\references
     {\vfill\eject
     {\noindent \bf References\par}
      \parindent=0pt
      \bigskip}

%

%
\def\refjl#1#2#3#4
   {\hangindent=16pt
    \hangafter=1
    \rm #1
   {\frenchspacing\sl #2
    \bf #3}
    #4\par}

%
\def\refbk#1#2#3
   {\hangindent=16pt
    \hangafter=1
    \rm #1
   {\frenchspacing\sl #2}
    #3\par}

%
\def\numrefjl#1#2#3#4#5
   {\parindent=40pt
    \hang
    \noindent
    \rm {\hbox to 30truept{\hss #1\quad}}#2
   {\frenchspacing\sl #3\/
    \bf #4}
    #5\par\parindent=16pt}

%
\def\numrefbk#1#2#3#4
   {\parindent=40pt
    \hang
    \noindent
    \rm {\hbox to 30truept{\hss #1\quad}}#2
   {\frenchspacing\sl #3\/}
    #4\par\parindent=16pt}

%

\def\ref#1{\noindent \hbox to 21pt{\hss
#1\quad}\frenchspacing\ignorespaces}

%
\def\frac#1#2{{#1 \over #2}}

%

%

%


\chardef\ii="10

%

%
\def\etal{{\sl et al\/}\ }

%

\catcode`\@=11
\def\vfootnote#1{\insert\footins\bgroup
    \interlinepenalty=\interfootnotelinepenalty
    \splittopskip=\ht\strutbox 
    \splitmaxdepth=\dp\strutbox \floatingpenalty=20000
    \leftskip=0pt \rightskip=0pt \spaceskip=0pt \xspaceskip=0pt
    \noindent\eightpoint\rm #1\ \ignorespaces\footstrut\futurelet\next\fo@t}

%
%
%
\catcode`\@=12
%
%



%
%





\def\NT{Nanotechnology}

%
%

\def\JCP{J. Chem. Phys.}

\def\JMP{J. Math. Phys.}

\def\NC{Nuovo Cim.}

\def\NP{Nucl. Phys.}
\def\PL{Phys. Lett.}
\def\PR{Phys. Rev.}
\def\PRL{Phys. Rev. Lett.}

\def\ZP{Z. Phys.}
%
%

%
%

%

%

%
\def\ns{\noalign{\vskip-3pt}}
%
\def\gap{\;\lower3pt\hbox{$\buildrel > \over \sim$}\;}
%
%
\def\lap{\;\lower3pt\hbox{$\buildrel < \over \sim$}\;}
\def\tqs{\hbox to 25pt{\hfil}}


%
%
%
\def\LaTeX{L\kern-.26em \raise.6ex\hbox{\fiverm A}%
   \kern-.15em\TeX}%
\def\AmSTeX{%
{$\cal{A}$}\kern-.1667em\lower.5ex\hbox{%
 $\cal{M}$}\kern-.125em{$\cal{S}$}-\TeX}

\endinput